\documentclass[prd, onecolumn, longbibliography, superscriptaddress, nofootinbib, floatfix, notitlepage]{revtex4-1}

\usepackage{amsmath}
\usepackage{graphicx}
\usepackage{dcolumn}
\usepackage{bm}
\usepackage{epsfig}
\usepackage{amssymb,latexsym,mathrsfs}
\usepackage{graphicx}
\usepackage{color}
\usepackage{xcolor}
\usepackage{ulem}
\usepackage{hyperref}

\hypersetup{
    colorlinks=true,
    linkcolor=red,
    citecolor=blue,
} 

\usepackage{subfigure}

\newcommand{\mpl}{M^2_{\rm pl}}

\usepackage{tikz}
\usepackage{mathrsfs} 
\usepackage{xfp} 
\usepackage[outline]{contour} 
\usetikzlibrary{decorations.markings,decorations.pathmorphing}
\usetikzlibrary{angles,quotes} 
\usetikzlibrary{arrows.meta} 
\contourlength{1.4pt}

\newcommand{\calI}{\mathscr{I}} 
\tikzset{>=latex} 
\colorlet{myred}{red!80!black}
\colorlet{myblue}{blue!80!black}
\colorlet{mygreen}{green!80!black}
\colorlet{mydarkred}{red!50!black}
\colorlet{mydarkblue}{blue!50!black}
\colorlet{mylightblue}{mydarkblue!6}
\colorlet{mypurple}{blue!40!red!80!black}
\colorlet{mydarkpurple}{blue!40!red!50!black}
\colorlet{mylightpurple}{mydarkpurple!80!red!6}
\colorlet{myorange}{orange!40!yellow!95!black}
\tikzstyle{cone}=[mydarkblue,line width=0.2,top color=blue!60!black!30,
                  bottom color=blue!60!black!50!red!30,shading angle=60,fill opacity=0.9]
\tikzstyle{cone back}=[mydarkblue,line width=0.1,dash pattern=on 1pt off 1pt]
\tikzstyle{world line}=[myblue!60,line width=0.4]
\tikzstyle{world line t}=[mypurple!60,line width=0.4]
\tikzstyle{particle}=[mygreen,line width=0.5]
\tikzstyle{photon}=[-{Latex[length=4,width=3]},myorange,line width=0.4,decorate,
                    decoration={snake,amplitude=0.9,segment length=4,post length=3.8}]
\tikzstyle{singularity}=[myred,line width=0.6,decorate,
                         decoration={zigzag,amplitude=2,segment length=6.17}]
\tikzset{declare function={%
  penrose(\x,\c)  = {\fpeval{2/pi*atan( (sqrt((1+tan(\x)^2)^2+4*\c*\c*tan(\x)^2)-1-tan(\x)^2) /(2*\c*tan(\x)^2) )}};%
  penroseu(\x,\t) = {\fpeval{atan(\x+\t)/pi+atan(\x-\t)/pi}};%
  penrosev(\x,\t) = {\fpeval{atan(\x+\t)/pi-atan(\x-\t)/pi}};%
  kruskal(\x,\c)  = {\fpeval{asin( \c*sin(2*\x) )*2/pi}};
}}
\def\tick#1#2{\draw[thick] (#1) ++ (#2:0.04) --++ (#2-180:0.08)}

\def\R{0.08} 
\def\e{0.08} 
\def\ang{45} 
\def\angb{acos(sqrt(\e)*sin(\ang))} 
\def\a{\R*sin(\ang)*sqrt(1-\e*sin(\ang)^2)/(1-\e*sin(\ang)^2)} 
\def\b{\R*sqrt(\e)*sin(\ang)*cos(\ang)/(1-\e*sin(\ang)^2)} 
\def\coneback#1{ 
  \draw[cone back] 
    (#1)++(-45:\R) arc({90-\angb}:{90+\angb}:{\a} and {\b});
  \draw[cone,shading angle=-60] 
    (#1)++(0,{\R*cos(\ang)/(1-\e*sin(\ang)^2)}) ellipse({\a} and {\b});
}
\def\conefront#1{ 
  \draw[cone] 
    (#1) --++ (45:\R) arc({\angb-90}:{-90-\angb}:{\a} and {\b})
     --++ (-45:2*\R) arc({90-\angb}:{-270+\angb}:{\a} and {\b}) -- cycle;
}

\begin{document}

\title{Tadpole Cosmology: Milne Solution as a Cosmological Constant Hideout} 

\author{Stephen Appleby}
\email{stephen.appleby@apctp.org}
\affiliation{Asia Pacific Center for Theoretical Physics, Pohang 37673, Korea} 
\affiliation{Department of Physics, POSTECH, Pohang 37673, Korea}

\author{Reginald Christian Bernardo}
\email{rbernardo@gate.sinica.edu.tw}
\affiliation{Institute of Physics, Academia Sinica, Taipei 11529, Taiwan}

\begin{abstract}
Dynamical cancellation frameworks present a potential means of mitigating the effect of a large vacuum energy, that would otherwise ruin the late-time, low energy dynamics of the Universe. Certain models in the literature, such as the Fab Four and Well Tempering, realize this idea by introducing some degeneracy in the dynamical equations. In this paper, we introduce a third potential route to self-tuning, and infer the existence of a new, exact Milne solution in the simplest tadpole plus cubic-Galileon scalar-tensor theory. We study the dynamics of the scalar field and metric in the vicinity of the Milne coordinate singularity, and find that the vacuum solution belongs to a more general family of Milne-like metrics. By numerically evolving the field equations for a range of initial conditions, we show that the Milne solution is not an attractor, and varying the initial scalar field data can lead to completely different asymptotic states; exponential growth of the scale factor, a static non-spatially flat metric or a severe finite-time instability in the scalar field and metric. We generalise the Milne solution to a class of FLRW spacetimes, finding that the tadpole-cubic Galileon model admits perfect-fluid-like solutions in the presence of matter. Finally, we present a second Horndeski model which also admits an exact Milne solution, hinting at the existence of a larger undiscovered model space containing vacuum-energy-screened solutions.
\end{abstract}

\date{\today} 

\maketitle

\section{Introduction}
\label{sec:introduction}

The Cosmological Constant (CC) problem is one of the outstanding issues in cosmology \cite{Weinberg:1988cp,Martin:2012bt, Padilla:2015aaa}. Particle physics places the `natural' scale of the vacuum energy anywhere between the Planck scale $\rho_{\rm vac} \sim {\cal O}(M_{\rm pl}^{4})$ and $\rho_{\rm vac} \sim {\cal O}({\rm GeV}^{4})$. Its precise value is sensitive to the ultraviolet (UV) cut-off of the effective field theory used to infer it. In General Relativity (GR), all forms of energy gravitate and the presence of a natural particle physics scale vacuum energy is completely inconsistent with the low energy state that we observe in the late time Universe. For the purposes of cosmology, the problem is resolved by introducing a `bare' cosmological constant into the Einstein-Hilbert action, and treating the vacuum energy as a free parameter to be fixed by observation. This is a perfectly legitimate resolution; from an effective field theory perspective the values of parameters such as $\rho_{\rm vac}$ are not predicted. However, the extreme discrepancy between the energy scale that is measured $\rho_{\rm vac} \sim {\cal O}(10^{-48} \ {\rm GeV}^{4})$ and what is naively expected $\rho_{\rm vac} \sim {\cal O}({\rm GeV}^{4})$ has led to a search for models that can naturally realise low energy vacuum states without fine tuning the vacuum energy. The problem is one of the most well-studied in the cosmology and high energy physics communities \cite{Carroll:2000fy, Garriga:2000cv,Csaki:2000wz, Forste:2000ft, Mukohyama:2003nw, Mukohyama:2003ac, Aghababaie:2003wz, Burgess:2011va, Carlip:2018zsk, Lombriser:2019jia,Wang:2019mbh, Lacombe:2022cbq, DAgostino:2022fcx, Bernardo:2022cck,Moreno-Pulido:2022phq, Belfiglio:2023eqi,Kaloper:2023xfl}. Attempts to tackle the problem typically involve renormalization, extra dimensions, quantum gravity corrections, unimodular gravity and more exotic approaches. To evade fine tuning, a powerful no-go theorem must be circumvented \cite{Weinberg:1988cp}.

To this end, an interesting scalar-tensor model was proposed in \cite{Charmousis:2011bf, Charmousis:2011ea}, in which empty Milne patches of spacetime were constructed despite the presence of an arbitrarily large vacuum energy. In this class of models, the vacuum energy is a free parameter, but low energy vacuum states are present for which it does not gravitate. Rather, the vacuum energy is dynamically cancelled by a time-evolving scalar field, which does not relax to a constant vacuum expectation value. This class of models provided an elegant mechanism to accommodate a natural particle physics scale cosmological constant. The approach relied on degeneracy in the field equations, which allowed for exact Milne solutions. Degeneracy in this case means that the scalar field equation identically vanishes when the Milne metric was imposed. A simple example of this approach is reviewed in Section \ref{sec:dynamicaltuningofthecc}, and various detailed aspects can be found in \cite{Copeland:2012qf, Appleby:2015ysa, Babichev:2015qma, Kaloper:2013vta, Copeland:2021czt,Appleby:2012rx}. In what follows, we call such dynamical tuning mechanisms `self-tuning'.  

Following on, a different degeneracy condition was proposed in \cite{Appleby:2018yci}, and was used in \cite{Appleby:2020dko} to find exact Minkowski space solutions in despite the presence of an arbitrarily large vacuum energy. In this class of so-called `Well Tempering' models, the simplest cubic Galileon plus tadpole Lagrangian was found to admit Minkowski space as the unique vacuum state for the metric, with an arbitrary vacuum energy not affecting the spacetime curvature `on-shell'\footnote{`On-shell' is terminology borrowed from \cite{Charmousis:2011bf, Charmousis:2011ea}, meaning that the metric is fixed exactly to its vacuum state, in this case Minkowski space.}. It was subsequently shown that these models admit FLRW cosmological solutions with epochs of matter and radiation, followed by an approach to Minkowski space as an asymptotic solution. This class of models has been studied within the context of observations \cite{Bernardo:2022vlj,Escamilla-Rivera:2023rop}, teleparallel extensions \cite{Bernardo:2021bsg, Bernardo:2021izq} and general cosmological  \cite{Appleby:2018yci,Appleby:2020dko,Emond:2018fvv,Appleby:2020njl, Linder:2020xey,Bernardo:2021hrz,Linder:2022iqi} solutions. Relaxing the degeneracy condition has been considered in \cite{Khan:2022bxs, Appleby:2022bxp, Linder:2022xms}. We provide a short review of this approach in Section \ref{sec:dynamicaltuningofthecc}.

The Milne slicing of Minkowski space is a useful testing ground for self-tuning models in the context of cosmology, because it possesses the symmetry properties of an FLRW metric and a (coordinate) singularity at $\tau = 0$. As we will elucidate below, self-tuning the vacuum energy using degeneracy turns the cosmological constant problem into an initial condition problem; the question becomes `what initial conditions for the scalar field and metric degrees of freedom yield a dynamical evolution that is consistent with observations, leading to a low energy, late time state?' This question is ambiguous in a global Minkowski spacetime, because there is no concept of a finite initial time-slice. In contrast, Milne possesses a unique point; $\tau = 0$, for which the metric possesses a coordinate singularity. This allows us to study the dynamical evolution of the fields from a special initial point to some $\tau \to \infty$ asymptotic state. The $\tau =0$ coordinate singularity will be promoted to a curvature singularity when matter is introduced, so performing an analysis in empty FLRW spacetime (Milne) can potentially provide some insight into how the singular point might be approached in a more realistic scenario. 

The derivation of the Fab Four model was definitive; the four coupled scalar field and graviton kinetic terms found in \cite{Charmousis:2011bf, Charmousis:2011ea} constitute the unique subset of Horndeski scalar-tensor Lagrangian terms which yield self-tuned Milne solutions, subject to certain conditions. In this work, we find a new class of self-tuned Milne and Milne-like spacetimes within the Horndeski model space. The existence of these solutions does not contradict the original Fab Four derivation -- in what follows we explain which assumptions are violated in arriving at these new Milne spacetimes. Our approach yields a third potential self tuning route, distinct and somehow weaker than Fab Four and Well Tempering. Interestingly, the original Well Tempering model of \cite{Appleby:2020dko} possesses an independent Milne solution, hence yielding two distinct scalar field foliations of Minkowski space.

The outline of this work is as follows. We introduce the Milne and Minkowski coordinate representations used throughout this work in Section \ref{sec:preamble}, and show how both were used in the Fab Four and Well Tempered Cosmological models for the dynamical cancellation of the CC in Section \ref{sec:dynamicaltuningofthecc}. Then, we show that the simplest Well Tempered tadpole--cubic Galileon model also admits a nontrivial Milne spacetime solution for which $\Lambda$ is dynamically cancelled, without relying on degeneracy. We fully explore this Milne solution and the more general family of dynamical spacetimes that it belongs to, and how the asymptotic final state of the metric and scalar field is sensitive to the initial configuration. We find new, exact FLRW solutions using the same methodology in Section \ref{sec:generalizationtoflrw}, showing that the Milne solution is a special case of a more general set of spacetimes for which the vacuum energy does not gravitate. In Section \ref{sec:othermilnesolutions}, we furthermore show by example that Milne solutions exist outside of the simplest tadpole--cubic Galileon Lagrangian, hinting at their possible existence in the larger non-Fab Four scalar-tensor theories. To conclude, we discuss further implications of our results on self tuning phenomenology and cosmology, and then give final remarks in Section \ref{sec:discussion}.

In a series of appendices we give some supporting details starting with an interesting graphical analogy between Milne space and the black hole interior (Appendix \ref{sec:analogy}). Then, we explicitly show the absence of a degeneracy condition in the new Milne solutions (Appendix \ref{sec:nonlinearity}) and derive the Minkowski solution in Milne coordinates and vice versa (Appendix \ref{sec:alternativeminkowski}). We lastly analytically derive the different late-time asymptotic states of the model considered in this work (Appendix \ref{sec:app_B}).

We work with the mostly plus metric signature $(-,+,+,+)$ and natural units $c = 1$. 

\section{Metrics}
\label{sec:preamble}

Throughout this work, we will focus on two different representations of Minkowski space, described using the following coordinate systems -- 

\begin{eqnarray}\label{eq:mink} & & ds^{2} = -dt^{2} + dr^{2} + r^{2} d\Omega^{2} , \\ 
\label{eq:milne} & & ds^{2} = -d\tau^{2} + \tau^{2} \left( {dy^{2} \over 1 + y^{2}} + y^{2} d\Omega^{2} \right)  ,
\end{eqnarray}

\noindent which we call Minkowski and Milne respectively in what follows -- despite the fact that both represent empty spacetime. The quantity $d\Omega^{2} = d\theta^{2} + \sin^{2}\theta d\phi^{2}$ is the standard covering of the unit two-sphere. 

The two coordinate systems are related via the transformations $\tau = \sqrt{t^{2} - r^{2}}$, $y = r/\sqrt{t^{2} - r^{2}}$ or $r = \tau y$, $t = \tau \sqrt{1+y^{2}}$. The `radial coordinate' $y$ is dimensionless, and the Milne metric only covers a patch of Minkowski space, with a coordinate singularity at $\tau = 0$. The Milne spacetime has identically vanishing curvature tensors, as the intrinsic curvature of the constant $\tau$ hypersurfaces exactly cancels the extrinsic curvature. The Milne metric can be considered as an empty FLRW spacetime with zero acceleration.

A notable feature of Milne space is that the coordinate singularity $\tau = 0$ cannot be reached by signals, corresponding to a light cone. One way to see this is by expressing the relation between the time coordinate in the two frames as $\tau^2 = t^2 - r^2$. However, the Milne patch of Minkowski only covers $t \geq r$ and time-like events and observers are restricted by  $\Delta t > \Delta r$. Thus, $\tau = 0$ cannot be probed from within the Milne patch. Appendix \ref{sec:analogy} presents an analogy between the Milne slicing of Minkowski space and a static black hole solution.

In what follows, we will generalise the Milne metric to the following form 

\begin{equation}\label{eq:gen_milne} ds^{2} = -d\tau^{2} + a^{2}(\tau) \left( {dy^{2} \over 1 - k y^{2}} + y^{2} d\Omega^{2} \right)  ,
\end{equation} 

\noindent with the scale factor $a(\tau)$ a generic function of $\tau$ and $k$ as a free parameter. For $a(\tau) = \tau$ and $k = -1$, or $a=1$ and $k=0$, the curvature tensors are identically zero. 

\section{Dynamical Tuning of the CC}
\label{sec:dynamicaltuningofthecc}

In this section, we briefly review two distinct yet related methods of dynamically cancelling the Cosmological Constant, to elucidate the importance of a so-called degeneracy condition.

\subsection{Fab Four Milne Solutions}

The original class of models, the Fab Four, possess exact Milne solutions for a class of scalar-tensor gravitational actions, by construction. For brevity, we only present one of the simplest models within the more general class, and direct the reader to \cite{Charmousis:2011bf,Charmousis:2011ea} for a complete classification. 

The Lagrangian that we consider is given by 

\begin{equation} {\cal L}_{\rm FF} = {M_{\rm pl}^{2}  \over 2}R + v_{0} G^{\mu\nu}\nabla_{\mu}\phi \nabla_{\nu}\phi - \Lambda , \end{equation} 

\noindent where $\Lambda$ is the bare cosmological constant, $G^{\mu\nu}$ is the Einstein tensor and $v_{0}$ is a free parameter of mass dimension $M^{-2}$. After constructing the covariant field equations for this Lagrangian, then inserting the metric (\ref{eq:gen_milne}), we arrive at the following equations -- 

\begin{eqnarray} \label{eq:fri_ff} & & 3M_{\rm pl}^{2} \left(  H^{2} + {k \over a^{2}}\right) = \Lambda  + 3 v_{0} \left(3 H^{2} + {k \over a^{2}} \right) (\phi')^{2}  , \\
\label{eq:da_ff} & & -M_{\rm pl}^{2}\left( 2H' + 3 H^{2} + {k \over a^{2}} \right) = -\Lambda  + v_{0} \phi' \left( \left[{k \over a^{2}} - 2H' - 3H^{2} \right] \phi' - 4 H \phi'' \right) , \\
\label{eq:sfe_ff} & & \left( H^{2} + {k \over a^{2}}\right)\phi'' +3 H^{3} \phi' + H \left(2H' + {k \over a^{2}}\right)\phi' = 0 ,
\end{eqnarray} 

\noindent where primes denote differentiation with respect to $\tau$ and $H = a'/a$. If we then fix the ansatz $a(\tau) = \tau$, $k = -1$ then the scalar field equation (\ref{eq:sfe_ff}) becomes identically satisfied, and the remaining equations are solved with 

\begin{equation}\label{eq:ff_m} \phi = \phi_{0} + {\sqrt{-\Lambda \over 24v_{0}}}\tau^{2} , \end{equation} 

\noindent where $\phi_{0}$ is a free parameter. The vacuum energy is dynamically cancelled by the scalar field in the Einstein equations for all $\tau$. This mechanism depends on some degeneracy in the field equations -- in this case the scalar field equation is trivially satisfied when the metric ansatz is imposed. This eliminates one of the dynamical equations once the metric has been fixed, and avoids the issue of the scalar field being over-constrained.

\subsection{Well-Tempered Minkowski Space Solutions}
\label{sec:WTMS}

Following from this idea, a different type of degeneracy was considered in \cite{Appleby:2020dko}. We again take the simplest example;

\begin{equation}\label{eq:wt} {\cal L}_{\rm WT} = {M_{\rm pl}^{2} \over 2} R + \epsilon X - {\epsilon^{2} \over \lambda^{3}} X \Box \phi - \lambda^{3}\phi - \Lambda   ,
\end{equation} 

\noindent where $X = -\nabla_{\nu}\phi\nabla^{\nu}\phi/2$ is the standard kinetic term of the scalar field $\phi$, $\epsilon$ is a free parameter and $\lambda$ is a mass scale (also free). This scalar-tensor Lagrangrian is invariant under the Galileon symmetry $\phi \to \phi + b_{\mu}x^{\mu} + c$ for constant $b_{\mu}$ and $c$. The covariant field equations are given by 

\begin{eqnarray} 
\nonumber & & {M_{\rm pl}^{2} \over 2} G_{\mu\nu}  - {\epsilon \over 2} g_{\mu\nu} X + {1 \over 2} g_{\mu\nu} \lambda^{3}\phi - {\epsilon \over 2} \phi_{\mu}\phi_{\nu} -  {\epsilon^{2} \over 2\lambda^{3}} \left[ \phi_{\mu} \phi_{\nu} \Box\phi - \phi_{\mu}\phi^{\lambda}\phi_{\nu\lambda} - \phi_{\nu}\phi^{\lambda}\phi_{\mu\lambda}  + g_{\mu\nu} \phi^{\lambda}\phi^{\beta} \phi_{\lambda\beta}\right]   + {1 \over 2} g_{\mu\nu} \Lambda = 0 , \\
\label{eq:eins} & &  \\
\label{eq:sfe} & & \epsilon \Box \phi  - \lambda^{3}  -  {\epsilon^{2} \over \lambda^{3}} \left[ \phi_{\alpha\beta}\phi^{\alpha\beta} + \phi^{\alpha}\Box \phi_{\alpha} - \phi^{\alpha}\nabla_{\alpha}\Box\phi  - (\Box\phi)^{2} \right]   = 0 \,, 
\end{eqnarray} 

\noindent and if we insert the metric ansatz (\ref{eq:mink}), $g_{\mu\nu} = \eta_{\mu\nu}$, and $\phi = \phi(t)$ into the field equations, they reduce to 

\begin{eqnarray} \label{eq:wt1} & & \Lambda + {\epsilon \over 2} \dot{\phi}^{2} + \lambda^{3}\phi = 0 , \\
\label{eq:wt2} & & {\epsilon^{2} \over \lambda^{3}} \dot{\phi}^{2} \ddot{\phi} + \epsilon \dot{\phi}^{2} = 0 , \\ 
\label{eq:wt3} & & \epsilon \ddot{\phi} + \lambda^{3} = 0 , 
\end{eqnarray} 

\noindent where dots denote differentiation with respect to $t$. Degeneracy is now realised in a different way -- the scalar field equation (\ref{eq:wt3}) is equivalent to the dynamical Einstein equation (\ref{eq:wt2}) once we have imposed the metric ansatz as an exact solution\footnote{Equation (\ref{eq:wt2}) has a second solution; $\dot{\phi} = 0$, but this does not solve the scalar field equation.}. As in the Fab Four, we have reduced the number of field equations, hence an exact solution exists once the metric ansatz has been imposed and the scalar field is not over-constrained. The solution is given by 

\begin{equation}\label{eq:wtmink} \phi(t) = \phi_{0} + \phi_{1} t - {\lambda^{3} t^{2} \over 2\epsilon} , \end{equation} 

\noindent where $\phi_{0}$, $\phi_{1}$ are integration constants, partially fixed by solving the Friedmann equation on some initial timeslice -- 

\begin{equation} \Lambda + {\epsilon \over 2} \phi_{1}^{2} + \lambda^{3}\phi_{0} = 0 \,.
\end{equation} 

\noindent The CC $\Lambda$ is dynamically cancelled at all subsequent times. The metric and the scalar field solution are invariant under time translations $t \to t + c$, which acts to redefine the constants $\phi_{0}$, $\phi_{1}$. 

Both of the examples considered above relied on reducing the number of equations once the metric has been fixed to a vacuum state. By imposing the Milne or Minkowski metric as an exact solution, one is left with a system of two dynamical equations and one constraint equation that $\phi(t)$ has to solve. This generically over-constrains the system, indicating that the initial ansatz is incorrect. These examples show that there is a subset of Horndeski models for which the ansatz can be imposed and a solution found, because the ansatz automatically solves one of the dynamical equations (or a linear combination thereof).

\section{Cubic Galileon and Milne}
\label{sec:cubicgalileonandmilne}
With the above discussion in mind, we now present our main findings in this section, that is, the existence of a Milne solution in the cubic Galileon (Section \ref{subsec:milne}). We then fully characterize the spacetimes that are causally adjacent to this Milne solution through asymptotic expansion (Section \ref{subsec:asymptoticsolutions}) and numerical analysis (Section \ref{subsec:numericalsolutions}).

\subsection{The Milne solution}
\label{subsec:milne}

We now search for Milne-like solutions to the Well-Tempering model given in the previous sub-section. Specifically, we insert the metric (\ref{eq:gen_milne}) into the field equations (\ref{eq:eins}, \ref{eq:sfe}), which yields the following field equations 

\begin{eqnarray}\label{eq:milne_sfe} & &  \left[ \epsilon - {6 \epsilon^{2}H \phi' \over \lambda^{3}} \right]\phi'' + \lambda^{3} + 3\epsilon H \phi' - 9{\epsilon^{2}  \over \lambda^{3}}H^{2}(\phi')^{2} - 3{\epsilon^{2} \over \lambda^{3}}H' (\phi')^{2} = 0 , \\
\label{eq:milne_da} & & M_{\rm pl}^{2} \left[ -2 H'   + {2k \over a^{2}}  \right] = \epsilon (\phi')^{2} + {\epsilon^{2} \over \lambda^{3}} (\phi')^{2} \phi'' - 3 {\epsilon^{2} \over \lambda^{3}} H (\phi')^{3} , \\ 
\label{eq:milne_fri} & & M_{\rm pl}^{2}\left[-3H^{2} - {3k \over a^{2}} \right] = -{\epsilon \over 2} (\phi')^{2} - \lambda^{3}\phi - \Lambda + {3 \epsilon^{2} \over \lambda^{3}}H (\phi')^{3} , 
\end{eqnarray} 

\noindent where we have also taken $\phi = \phi(\tau)$. 

To obtain the Minkowski solution in the previous sub-section, we fixed the metric completely in the field equations to $g_{\mu\nu} = \eta_{\mu\nu}$, and found that the two non-trivial dynamical equations are equivalent after this imposition. Following this idea, we impose the Milne spacetime $a = \tau$, $k=-1$ so $H^{2} = \tau^{-2}$ exactly in the above equations, which become;

\begin{eqnarray} \label{eq:hub1} & & 0  = \Lambda + \epsilon {(\phi')^{2} \over 2} - {3\epsilon^{2} (\phi')^{3} \over \lambda^{3} \tau}   + \lambda^{3} \phi , \\ 
\label{eq:dhm1} & & 0 =  {\epsilon^{2} \over \lambda^{3}}(\phi')^{2}\phi'' - {3\epsilon^{2}(\phi')^{3} \over \lambda^{3} \tau}    +  \epsilon (\phi')^{2}  ,  \\
\label{eq:sfem} & & 0 =   \left[ \epsilon - {6 \epsilon^{2} \phi' \over \lambda^{3} \tau} \right] \phi'' + {3 \epsilon \over \tau}\phi'  + \lambda^{3} - {6 \epsilon^{2}   \over \lambda^{3} \tau^{2}} (\phi')^{2} \,.
\end{eqnarray}  

Different to the previous Well Tempered and Fab Four cases, the dynamical equations (\ref{eq:dhm1}) and (\ref{eq:sfem}) are not equivalent, and neither of them is trivially satisfied after imposing the metric ansatz. This indicates that degeneracy is not realised. We explicitly show that the equations are not equivalent in Appendix \ref{sec:nonlinearity}. However despite this, the system of equations do admit an exact solution of the form 

\begin{equation}\label{eq:mils} \phi(\tau) = \phi_{0} + {\lambda^{3} \tau^{2} \over 4\epsilon} , \end{equation} 

\noindent where the Hamiltonian reduces to 

\begin{equation}\label{eq:tad_m} \Lambda + \lambda^{3} \phi_{0} = 0 , \end{equation} 

\noindent and $\phi_{0}$ is an arbitrary constant in principle, but in practice is fixed via the Hamiltonian constraint at $\tau = 0$. This partially defines the `initial condition' of $\phi$, although we return to this point shortly. 

Similarly to the self-tuning models of the previous sections, we have inserted a metric ansatz and then inferred an exact solution for the scalar field, but this time in the absence of any degeneracy condition. This is a curious state of affairs, because the imposition of an exact metric ansatz would typically over-constrain the scalar field dynamics. Based on this example, we see that there is a third potential mechanism for self-tuning -- the scalar field and Einstein equations (\ref{eq:dhm1}) and (\ref{eq:sfem}) can admit the same solution when a metric ansatz is imposed even though they are distinct equations. The differential equations do not need to be identical, nor does one need to vanish. They simply have to admit the same solution. Non-linear equations such as these typically admit large solution spaces, and they can partially overlap. 

However, an important distinction is that the solution above does not have the same generality as the Milne solution (\ref{eq:ff_m}) in the Fab Four model, in which the scalar field has a single free parameter $\phi_{0}$ that is not fixed by the Friedmann equation. Conversely, $\phi_{0}$ must satisfy (\ref{eq:tad_m}) in this section. In the Fab Four model, one equation is eliminated after insertion of the metric, leaving a single second order dynamical equation for $\phi(\tau)$ which requires two pieces of initial data. In contrast, the solution space of the two dynamical equations (\ref{eq:dhm1},\ref{eq:sfem}) only partially overlap -- each equation has a separate set of solutions and only the subset (\ref{eq:mils}) solves them simultaneously. This indicates the presence of additional solutions beyond Milne. The approach considered here may therefore be considered as distinct and somehow weaker than the Fab Four and Well Tempering, both of which require two pieces of initial data for $\phi$, $\phi'$ after insertion of the metric ansatz, and are only partially fixed by the Hamiltonian constraint\footnote{The requirement of two initial pieces of data is not immediately clear from the Fab Four solution (\ref{eq:ff_m}), which only contains $\phi_{0}$. The initial condition fixes $\phi'=0$ at $\tau =0$, to avoid spurious singularities in the scalar field energy density. Anyway, after the application of the Hamiltonian constraint on some initial timeslice, both Fab Four and Well Tempering possess a single residual degree of freedom.}. The existence of these new Milne solutions does not contradict the original derivation of the Fab Four, which was predicated on degeneracy and found a unique set of four scalar field kinetic terms such that the scalar field equation vanishes when evaluated at the vacuum. The additional requirement in \cite{Charmousis:2011bf, Charmousis:2011ea} that the Milne solution is preserved through a phase transition implies that the solution is an attractor, and as we will show in the following section the new Milne solutions found here are not. 

If we keep $a(\tau)$ in the metric (\ref{eq:gen_milne}) as a free function, as opposed to fixing it to Milne, then we have a pair of non-linear second order differential equations for $a(\tau)$ and $\phi(\tau)$. To solve this system, we require four initial conditions $a_{i}$, $a'_{i}$, $\phi_{i}$ and $\phi'_{i}$ on some initial time slice $\tau = \tau_{i}$. The Friedmann equation must be solved initially, which fixes a single initial condition, but this still leaves three pieces of initial data unspecified. The method that we used to infer a solution was to impose the form of the metric exactly, and arrived at a solution containing one free parameter $\phi_{0}$, which is then fixed via the Friedmann equation. This approach has obfuscated the more general solution space. Understanding what has happened to the freedom to specify the initial data is further complicated by the fact that we selected the initial time slice to be $\tau = 0$, which is a coordinate singularity for the Milne metric. In such a case, the initial conditions are partially fixed by the demanding that the solution is regular. 

Since we are searching for spacetimes in which the cosmological constant is dynamically cancelled, an important question is the time domain over which we can construct such a solution. For an ansatz such as (\ref{eq:milne}), we must consider the range of $\tau$ over which the metric is free of singularities and a solution can be found. Then, we should consider the generality of initial conditions imposed on $a(\tau)$ and $\phi(\tau)$, and the conditions under which a Milne-like solution is dynamically approached. At the same time, we should understand that the Universe is not a Milne spacetime, because it is not empty. When we include matter/radiation, the Milne coordinate singularity at $\tau = 0$ is promoted to a curvature singularity, and it is inappropriate to select initial conditions at this point. For any realistic attempt to cancel the vacuum energy, we must incorporate some model of inflation and embed the self-tuning Galileon into it, or even propose a model of inflation based on the Galileon. The point is, we have shifted the cosmological constant problem into an initial condition problem, and the fine tuning argument becomes `how likely are the initial conditions for $H$, $\phi$ and $\phi'$, which lead to an expansion rate that is consistent with our observations when we include the Galileon dynamics over the entire dynamical history?' Answering this question is beyond the scope of the current work. However, for the Milne spacetime solution just found, we can make progress by studying the dynamics of the field $\phi(\tau)$ and $H(\tau)$ away from the Milne solution. Away from the `on-shell' dynamics, the spacetime will be sensitive to the vacuum energy $\Lambda$, but not in the same way as in GR. This is true for models that seek to dynamically cancel the vacuum energy.  

In GR, there are no such issues -- we have a single field (the scale factor), a dynamical equation which requires two pieces of initial data and a constraint equation which fixes one. The solution that satisfies both equations is $a(\tau) = \tau + \tau_{0}$, and $\tau$ can always be redefined such that $a(\tau = 0) = 0$. In Galileon Milne, we have two fields (scale factor and scalar field), require four initial data and only have a single constraint to fix one. This leads to a much broader family of solutions to be explored.

\subsection{Series Solution Around $\tau = 0$}
\label{subsec:asymptoticsolutions}

With the above discussion in mind, we will focus on the metric ansatz (\ref{eq:gen_milne}), fix $k=-1$ and keep $a(\tau)$ as free. We consider the region $\tau \geq 0$, and initially expand the equations around $\tau = 0$. This expansion behaves very differently depending on our choice of initial condition for $a(\tau)$; either $a(\tau=0) = 0$ or $a(\tau=0) = {\rm const} > 0$. We focus predominantly on the case $a(\tau = 0) = 0$, since the introduction of matter will generate this singular initial condition naturally. There is a large solution space to this system of equations, but we are only interested in the subset that has some relevance to cosmology. 

The field equations for this system are given in (\ref{eq:milne_sfe}-\ref{eq:milne_fri}). Searching for a more general set of solutions beyond Milne, we expand $a(\tau)$ and $\phi(\tau)$ around $\tau = 0$ as

\begin{eqnarray}
\label{eq:aexp1} a(\tau) &=& a_{1}\tau + {1 \over 2!} a_{2}\tau^{2} + {1 \over 3!} a_{3} \tau^{3} + {1 \over 4!} a_{4}\tau^{4} + {1 \over 5!} a_{5}\tau^{5} + {\cal O}(\tau^{6}) \\
\label{eq:phiexp1} \phi(\tau) &=& \phi_{0} + \phi_{1}\tau + {1 \over 2!} \phi_{2}\tau^{2} + {1 \over 3!} \phi_{3}\tau^{3}  + {1 \over 4!} \phi_{4}\tau^{4} + {1 \over 5!} \phi_{5}\tau^{5} + {\cal O}(\tau^{6})
\end{eqnarray}

\noindent where we have imposed one initial condition $a(\tau=0)=0$ with this expansion ansatz. We have also assumed that an analytic solution exists around $\tau = 0$. Inserting this expansion into the field equations, and solving order by order, we infer that there are two branches: 

\begin{eqnarray}
\label{eq:a_branch1} a(\tau) &=& \tau + \left( \dfrac{\Lambda + \lambda^3 \phi_0}{3 \mpl} \right) \dfrac{\tau^{3}}{3!} + \left( \dfrac{\Lambda + \lambda^3 \phi_0}{3 \mpl} \right)^2 \dfrac{\tau^{5}}{5!}  + {\cal O}\left(\tau^{7}\right) , \\
\label{eq:phi_branch1} \phi(\tau) &=& \phi_{0} + \dfrac{\lambda^3}{2\epsilon} \dfrac{\tau^{2}}{2!} - \dfrac{7\lambda^3}{8 \epsilon} \left( {\Lambda + \lambda^3 \phi_0 \over 3M_{\rm pl}^{2}}\right) \dfrac{\tau^{4}}{4!} + {\cal O}\left(\tau^{6}\right) ,
\end{eqnarray}

and 

\begin{eqnarray}
\label{eq:a_branch2} a(\tau) &=& \tau + \left( \dfrac{\Lambda + \lambda^3 \phi_0}{3 \mpl} \right) \dfrac{\tau^{3}}{3!} + \left[ \left( \dfrac{\Lambda + \lambda^3 \phi_0}{3 \mpl} \right)^2 - \dfrac{2 \lambda^6}{9 \mpl \epsilon} \right] \dfrac{\tau^{5}}{5!}  + {\cal O}\left(\tau^{7}\right) , \\
\label{eq:phi_branch2} \phi(\tau) &=& \phi_{0} - \dfrac{\lambda^3}{6\epsilon} \dfrac{\tau^{2}}{2!} + \dfrac{5\lambda^3}{24 \epsilon} \left( {\Lambda + \lambda^3 \phi_0 \over 3M_{\rm pl}^{2}} \right) \dfrac{\tau^{4}}{4!}  + {\cal O}\left(\tau^{6}\right) ,
\end{eqnarray}

\noindent which we refer to as branch I and II in what follows, respectively. This expansion partially answers the question regarding initial data -- we are free to impose $\phi_{0}$, $\phi_{1}$, $a_{0}$, $a_{1}$. We fixed $a_{0}=0$, since we are investigating solutions for which $a(\tau) \to \tau \to 0$ at $\tau =0$, anticipating that this form is relevant for cosmology. However, this fixes $a_{1} = 1$, and also $\phi_{1} = 0$ to remove singular terms in the expansion at $\tau = 0$. Demanding $a_{0} = 0$, forces us to select these initial conditions. Then, $a_{2} = 0$ solves the dynamical equations and the Friedmann equation at the first non-singular order. The next order reads

\begin{equation} \Lambda + \lambda^{3} \phi_{0} - 3 a_{3} M_{\rm pl}^{2} = 0 \,.
\end{equation} 

\noindent We are free to specify $\phi_{0}$, and specifically the Friedmann equation does not demand that we select $\lambda^{3}\phi_{0} + \Lambda = 0$ as in the Milne solution.  

An interesting consequence of the non-linearity of the equations is the existence of two branches of solutions, which are explicitly presented above. The first branch contains the Milne solution found previously, with $R = 0$ and $a_i = 0$ for $i \geq 2$ (flat spacetime) if we fix $\phi_{0} = -\Lambda/\lambda^{3}$. However, more generally we do not have to make this choice, and if we do not then both $a_{i} \neq 0$, $\phi_{i} \neq 0$ for $i \geq 2$ and subsequently $R \neq 0$. The second branch, which has $\phi \sim \phi_0 - \lambda^3 \tau^2/12\epsilon$, is a non-empty spacetime with $a_5 \neq 0$ even if we impose the initial condition $\Lambda + \lambda^3 \phi_0 = 0$. For this solution, the Ricci scalar is non-zero for $\tau > 0$ and cannot be made to be identically zero for any choice of $\phi_{0}$. For both branches, departures from the exact Milne solution $a(\tau) = \tau$ are suppressed by the Planck mass via the dimensionless combination $\sim (\Lambda + \lambda^{3}\phi_{0})\tau^{2}/M_{\rm pl}^{2}$.

Both branches of solutions have the same initial condition at $\tau = 0$; specifically $\phi'(\tau=0) = 0$, $a(\tau=0) = 0$ and $a'(\tau=0) =1$, which indicates that we cannot uniquely determine the dynamical evolution from $\tau =0$. This is an additional complication not present for the GR Milne solution, raising the question of whether the Galileon Milne solution is extendable to $\tau < 0$. Regardless, the point $\tau = 0$ will also not be suitable as an initial time-slice when matter is introduced, due to the presence of a curvature divergence at this point. For the present case, any choice of initial time $\tau_{i} >  0$ will allow us to distinguish the two branches of solutions. For some non-zero $\tau_{i}$, the two branches are separated by the approach to zero $\phi' \to 0^{\pm}$ as $\tau \to 0$. The branches do not intersect at any $\tau > 0$ and $\phi'$ remains either strictly positive or negative depending on the branch. In what follows, we refer to these expansion solutions as `Milne-like'. 

The $\tau =0$ point is a coordinate singularity for the Milne spacetime, and it is straightforward to show that it is also a coordinate singularity for all `Milne-like' generalisations considered in this section. The Ricci and Kretschmann scalars can be expanded as
\begin{equation}
    R = \frac{36 a_2}{\tau } +6 \left(12 a_3-6 a_2^2\right) + {\cal O}\left(\tau^1\right) ,
\end{equation}
and
\begin{equation}
    K = \frac{240 a_2^2}{\tau ^2}+\frac{48 \left(18 a_2 a_3-10 a_2^3\right)}{\tau }+12 \left(60 a_2^4-168 a_3 a_2^2+112 a_4 a_2+72 a_3^2\right) + {\cal O}\left(\tau^1\right) \,.
\end{equation}

\noindent The coefficient $a_{2}$ determines the regularity of the curvature invariants, and is zero for both branches I and II.

If we insert the same $\tau$ expansion of $a(\tau)$ and $\phi(\tau)$ around $\tau = 0$ for the Fab Four model, there are no branching solutions. There is only one solution, because at each order the scalar field equation fixes one of the parameters; typically $a_{i} = 0$, $i>1$, and then the Einstein equation fixes the corresponding $\phi_{j} = 0$ for $j > 2$. This confirms that in Fab Four the Milne metric encompasses an entire family of solutions, characterised by the free parameter $\phi_{0}$, whereas for the tadpole Galileon we can say that there exists an exact Milne spacetime, but it belongs to a more general family. This family can be considered as an explicit realisation of the Milne-like spacetimes in \cite{Ling:2018tih}, which also arrived at the conclusion that the Cosmological Constant can be related to the initial conditions of a dynamical system and specifically to $a'''$ (see also \cite{Ling:2017uxe,Ling:2018jzl,Klinkhamer:2019mti,Klinkhamer:2019rio,Ling:2022kjk} for a detailed discussion of the Milne spacetime and its relevance to cosmology).

\subsection{Numerical Solutions}
\label{subsec:numericalsolutions}

\begin{figure}[h!]
 \begin{center}
  \includegraphics[width=0.95\textwidth]{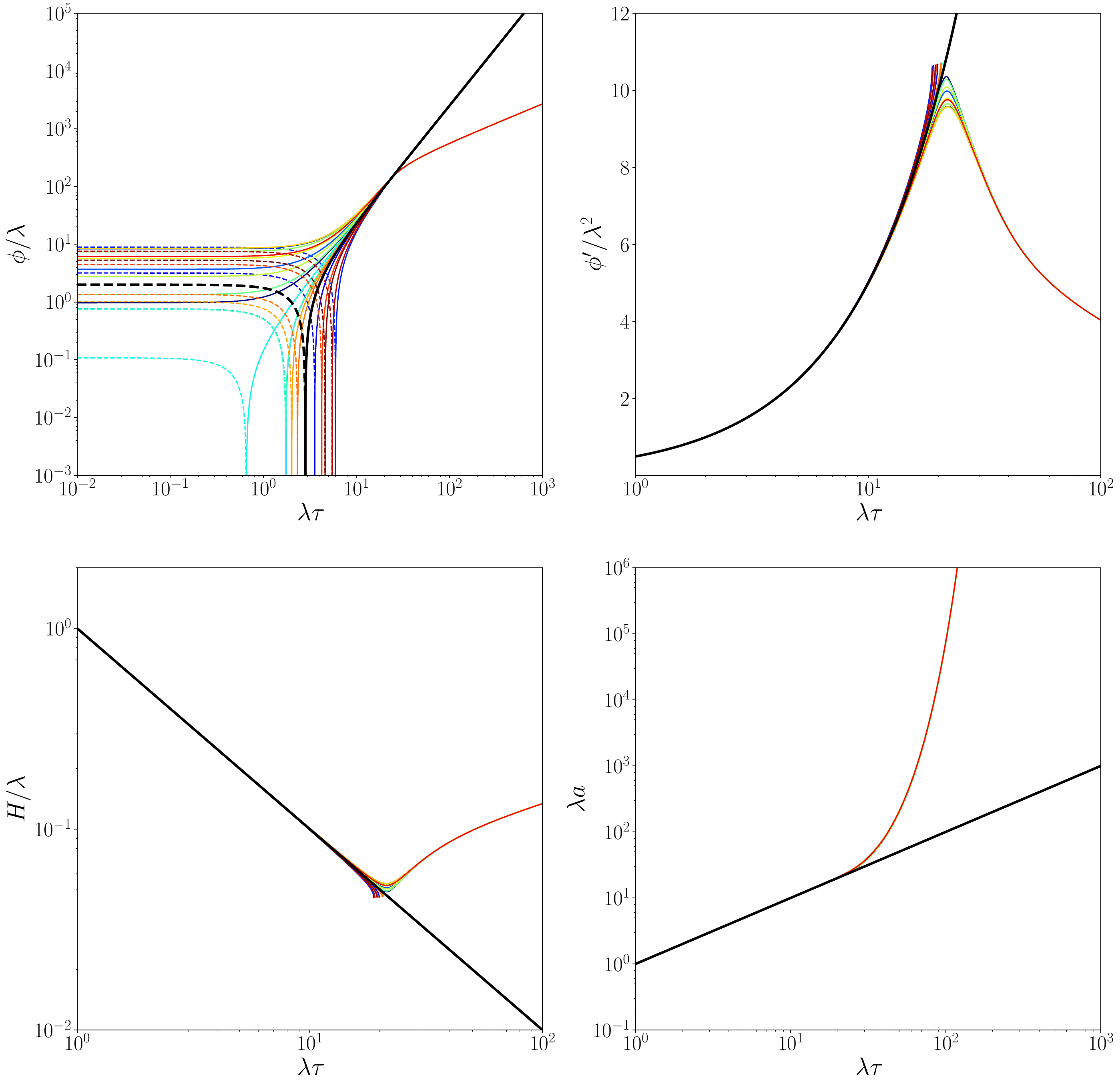}
 \end{center}
 \caption{\label{fig:1} Numerical solutions to \eqref{eq:milne_sfe}, \eqref{eq:milne_da}, and \eqref{eq:milne_fri} provided branch I initial conditions (\eqref{eq:a_branch1} and \eqref{eq:phi_branch1}) with $\lambda = 1$, $\Lambda = 10^{-2}\lambda^{4}$, and $\phi_0 = -x \Lambda/\lambda^3$ where $x$ is a randomly chosen in the range $-10^{-3} \leq x \leq 10^3$. The thick solid black line corresponds to $x = 1$; i.e. the special case $\Lambda + \lambda^3 \phi_0 = 0$. For the other colored curves, solid (dashed) curves represent positive (negative) numbers.
}
\end{figure}

\begin{figure}[h!]
 \begin{center}
  \includegraphics[width=0.95\textwidth]{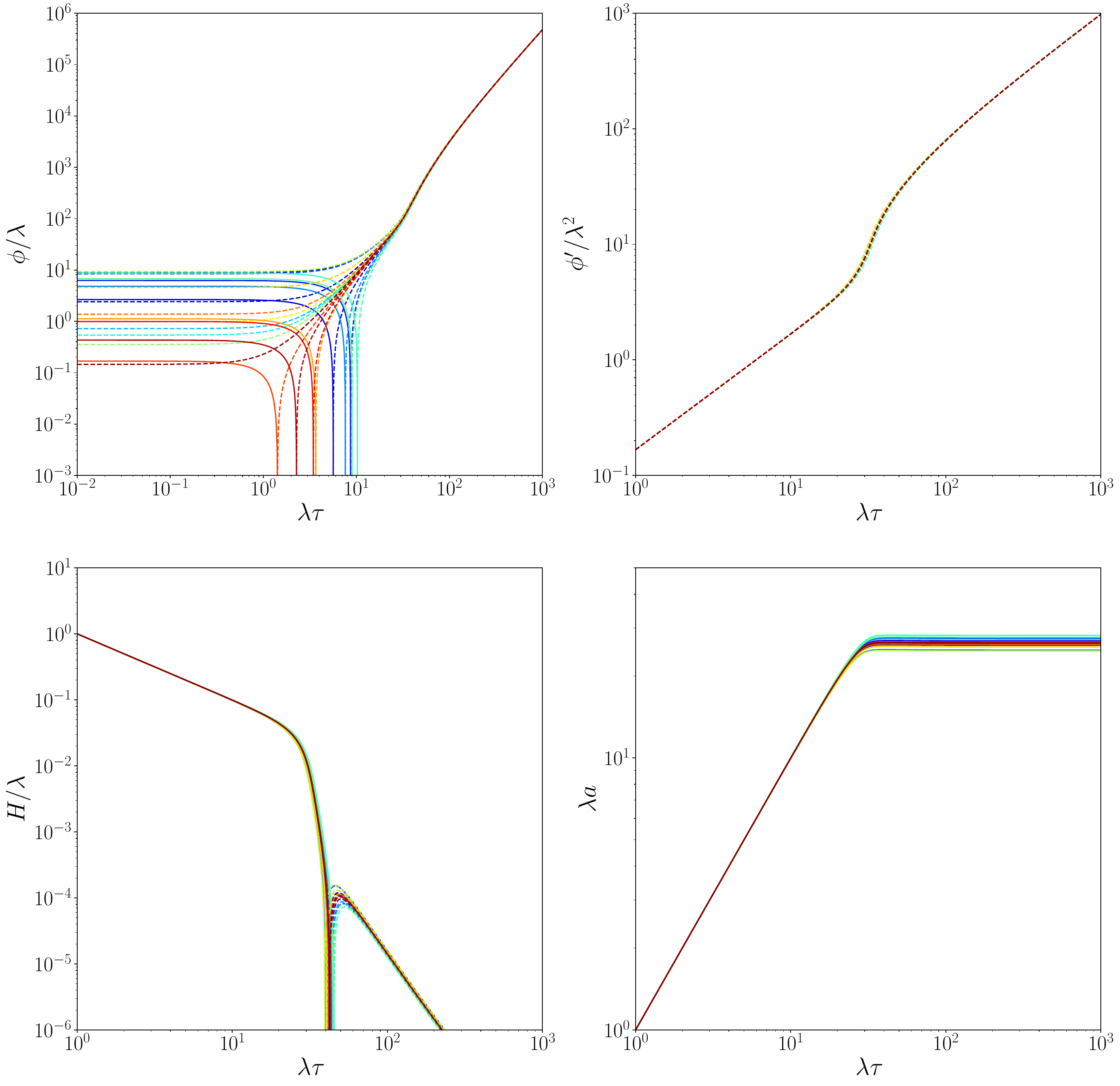}
 \end{center}
 \caption{\label{fig:3} Numerical solutions to \eqref{eq:milne_sfe}, \eqref{eq:milne_da}, and \eqref{eq:milne_fri} provided branch II initial conditions (\eqref{eq:a_branch2} and \eqref{eq:phi_branch2}) with $\lambda = 1$, $\Lambda = 10^{-2}\lambda^{4}$, and $\phi_0 = -x \Lambda/\lambda^3$ where $x$ is a randomly chosen in the range $-10^{-3} \leq x \leq 10^3$. This branch reveals a static, non-spatially flat final state. 
}
\end{figure}

\begin{figure}[h!]
 \begin{center}
  \includegraphics[width=0.45\textwidth]{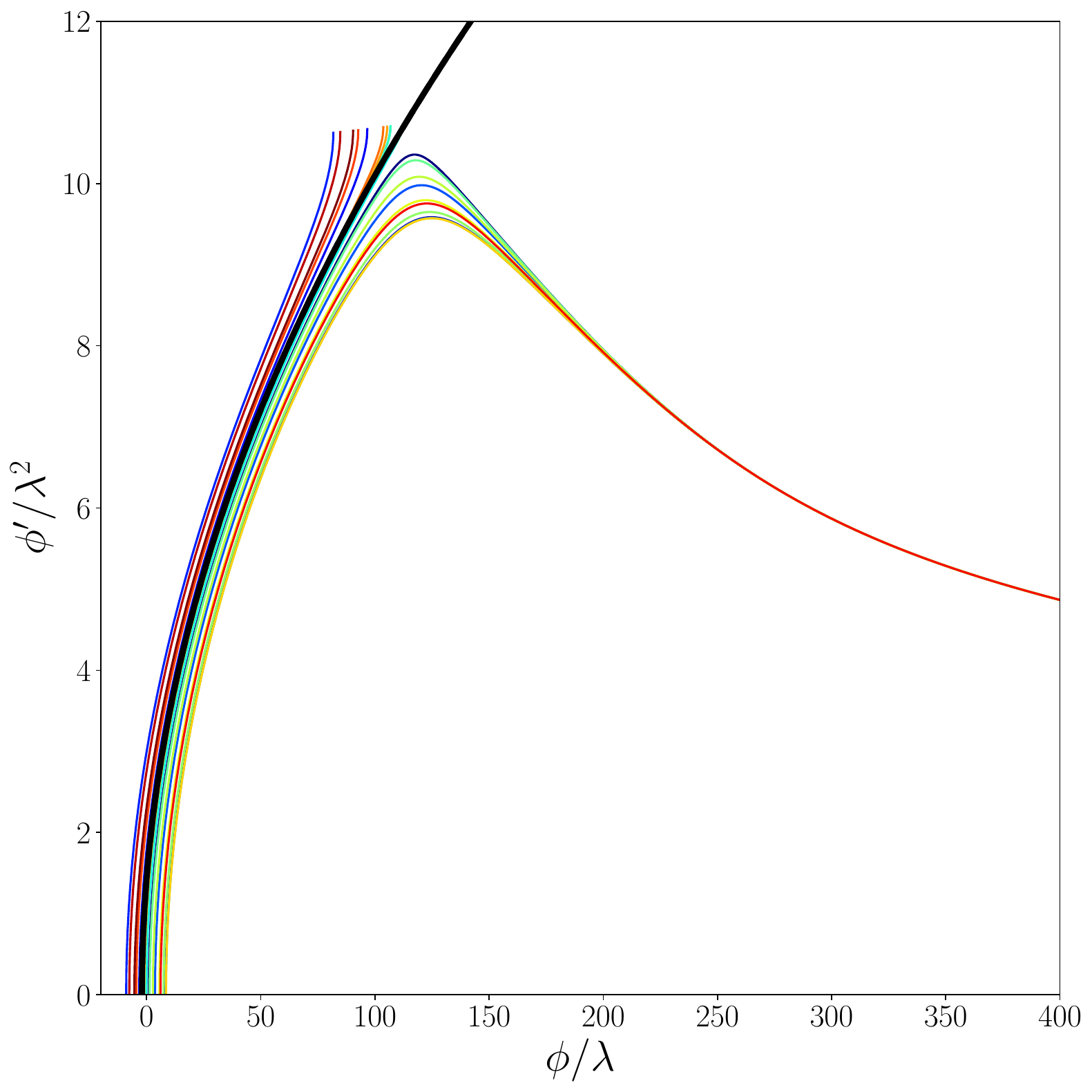}
  \includegraphics[width=0.45\textwidth]{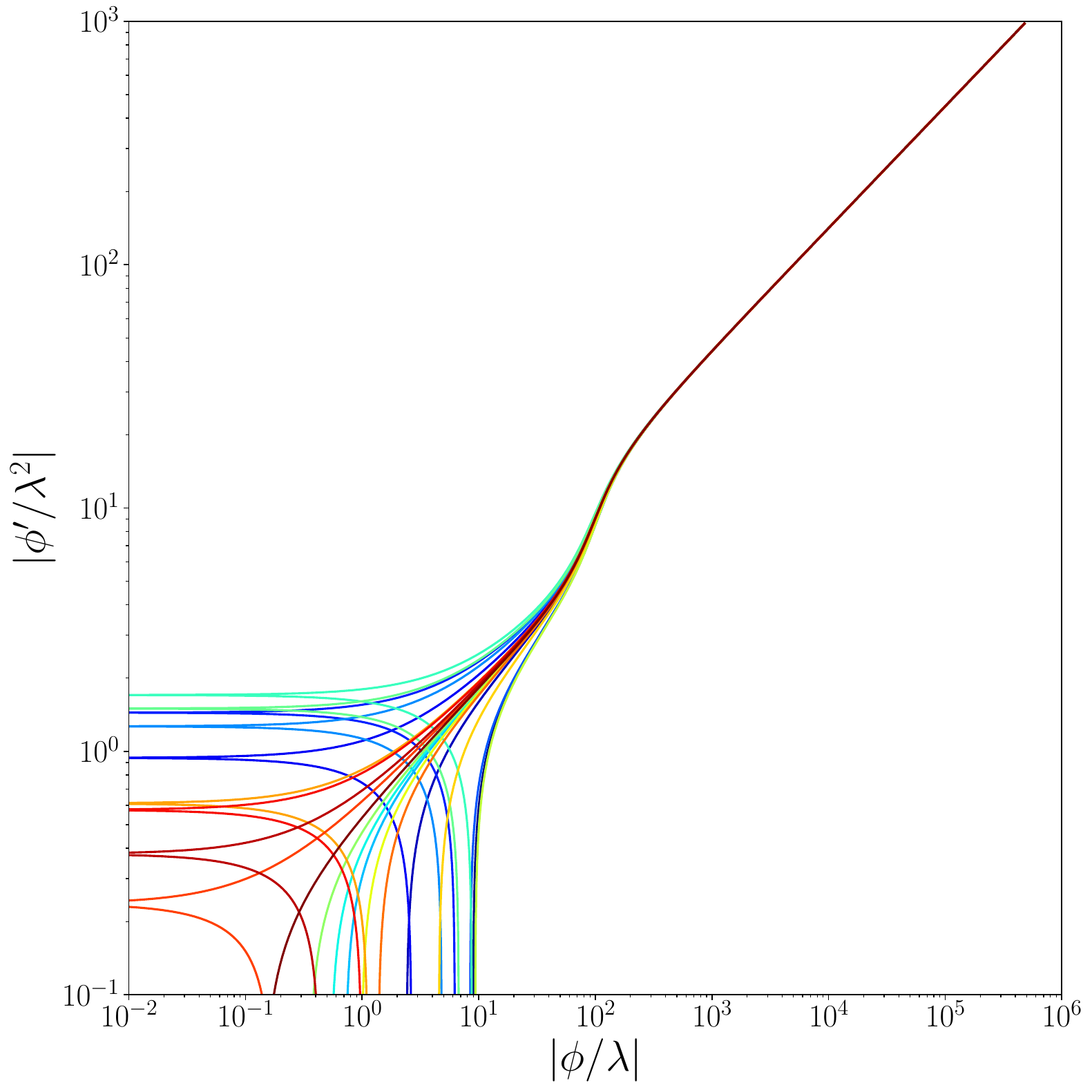}
 \end{center}
 \caption{\label{fig:2} Phase space representation of the numerical solutions to \eqref{eq:milne_sfe}, \eqref{eq:milne_da}, and \eqref{eq:milne_fri} provided [left] branch I (\eqref{eq:a_branch1} and \eqref{eq:phi_branch1}) and [right] branch II (\eqref{eq:a_branch2} and \eqref{eq:phi_branch2}) initial states with $\lambda = 1$, $\Lambda = 10^{-2}\lambda^{4}$, and $\phi_0 = -x \Lambda/\lambda^3$ where $x$ is a randomly chosen in the range $-10^{-3} \leq x \leq 10^3$. For the branch I solutions, the thick solid black line corresponds to $x = 1$; i.e. the special case $\Lambda + \lambda^3 \phi_0 = 0$.
}
\end{figure}

To go beyond the $\tau \simeq 0$ expansion, we must numerically evolve the equations. In this section we evolve equations (\ref{eq:milne_sfe},\ref{eq:milne_da}) for a range of scalar field initial conditions, solving the Friedmann equation (\ref{eq:milne_fri}) at some initial time and then using it as a consistency check that the dynamical equations are being correctly solved at each subsequent time-step. We fix $\lambda = 1$ and define all time and mass scales with respect to this choice. For some initial $\lambda \tau_{i} = 10^{-2}$, we use the expansions (\ref{eq:aexp1}, \ref{eq:phiexp1}) up to fifth order in $\tau_{i}$, fixing the $a_{0-5}$, $\phi_{0-5}$ parameters precisely according the the series solutions (\ref{eq:a_branch1}-\ref{eq:phi_branch2}), varying $\phi_{0}= -x\Lambda/\lambda^{3}$ where $x$ is selected randomly over the range $-10^{3} \leq x \leq 10^{3}$. We fix $\Lambda = 10^{-2}\lambda^{4}$ and $M_{\rm pl} = 10^{2}\lambda$; we are not introducing any large hierarchy between $\lambda$, $\Lambda$ and $M_{\rm pl}$. We repeat our analysis for both branches of solutions, characterised by $\phi_{2} = \lambda^{3}/2\epsilon$ and $\phi_{2} = -\lambda^{3}/6\epsilon$ in the expansion. In Figure \ref{fig:1} we present $\phi$, $\phi'$, $a$ and $H$ (from top left clockwise) for $N_{\rm r} = 20$ random initial conditions set by $\phi_{0}$ for branch I, and in Figure \ref{fig:3} $N_{\rm r} = 20$ different random initial conditions for branch II. Figure \ref{fig:2} depicts the corresponding phase space trajectories of these numerical solutions. In each panel of Figure \ref{fig:1} and the left panel of Figure \ref{fig:2}, the thick black line is the exact Milne solution with $\lambda^{3}\phi_{0} + \Lambda = 0$. Each random initial condition is represented by a different coloured track, and solid/dashed lines indicate the solution is positive/negative.

Branch I includes the exact Milne solution, but it is not an attractor if we vary the initial condition $\phi_{0}$. Specifically, from Figure \ref{fig:1} we find that if we set $\phi_{0} < -\Lambda/\lambda^{3}$ or $\phi_{0} > -\Lambda/\lambda^{3}$, we obtain distinct dynamical behaviour for both $\phi$ and $H$. By taking $\phi_{0} < -\Lambda/\lambda^{3}$, the scalar field and $H$ undergo an extreme instability at some finite time $\tau = \tau_{s}$. At this point, the highest derivatives $H'$ and $\phi''$ grow to extreme values, at which point our numerical scheme fails. This behaviour is presented via an abrupt end to the dynamical tracks in Figure \ref{fig:1}, to the left of the black line in the top right and bottom left panels. The end points correspond to the times at which the equations become numerically stiff. On approach to this point, the scale factor and scalar field remain perturbatively close to the Milne solution, but $H'$ and $\phi''$ grow arbitrarily. We study this behaviour analytically in Appendix \ref{sec:app_B}. This is not a sudden singularity of the kind studied in \cite{Nojiri:2005sx}, since the scale factor also exhibits (logarithmic) growth on approach to $\tau \to \tau_{\rm s}$. 

Conversely, if we set $\phi_{0} > -\Lambda/\lambda^{3}$ then the solution is regular over the timescales probed. The scalar field velocity $\phi'$ exhibits a turning point and starts to decrease (cf top right panel, Figure \ref{fig:1}). The scale factor undergoes exponential expansion after dallying in the initial Milne-like phase (cf. bottom right panel, Figure \ref{fig:1}). 

Based on the numerical analysis, any deviation from the exact Milne solution yields wildly different final states for both the expansion rate $H$ and the scalar field $\phi$. The dichotomy between the different end points in the dynamics -- finite time instability or exponential growth in $a$ -- is best observed in the phase space $(\phi,\phi')$, which is presented in the left panel of Figure \ref{fig:2}. These are the $N_{r} = 20$ dynamical tracks for $\phi$, $\phi'$ in Figure \ref{fig:1}, presented parametrically. The solid black line is again the Milne solution, which acts as a boundary separating the two generic final states. If we select $\Lambda + \lambda^{3}\phi_{0} < 0$, then both $\phi_{2}$ and $\phi_{4}$ are positive -- $\phi$ grows faster than in the Milne case. In contrast for $\Lambda + \lambda^{3}\phi_{0} > 0$ the higher derivative contribution $\phi_{4}$ is negative, which causes the maxima in the $(\phi',\phi)$ phase space (cf left panel, Figure \ref{fig:2}). If $\phi'$ grows too quickly, the finite time instability is approached. This corresponds to the point at which the coefficient of $\phi''$ in the field equations approaches zero:
\begin{equation} \epsilon - 6{\epsilon^{2} \over \lambda^{3}} H\phi' + {3 \over 2} {\epsilon^{4} \over M_{\rm pl}^{2} \lambda^{6}}(\phi')^{4} \to 0 \,.
\end{equation} 

Next we consider branch II, characterised by setting $\phi' < 0$ at $\tau = \tau_{i}$. This branch does not admit an exact Milne solution for any choice of initial condition; however we have shown that it does admit a Milne-like solution in the $\tau \simeq 0$ expansion. Evolving this branch numerically, we present the dynamics in Figure \ref{fig:3}. We see that  the Milne-like initial state is followed by a smooth transition to a regular asymptotic solution $\phi'' \to -\lambda^{3}/\epsilon$ and $a \to \alpha_{0}$ for some constant $\alpha_{0}$. The scalar field $\phi$ grows as $-\tau^{2}$ as $\tau \to \infty$, but all curvature invariants and energy densities remain finite. The $(\phi,\phi')$ phase space for this branch is presented in the right panel of Figure \ref{fig:2}. Branch II evolves according to $\phi' < 0$ as expected. In the bottom left panel of Figure \ref{fig:3} we see that $H(\tau) \simeq \tau^{-1}$ for a period, but crosses zero and then approaches zero from below: $H \to 0^{-}$ as $\tau \to \infty$. 

By studying the dynamical system numerically, we see that the Milne solution is a particular dynamical track that corresponds to a very particular initial condition. Allowing the initial scalar field value $\phi_{0}$ to vary, we find three generic final states beyond Milne. We present these three large $\tau$ asymptotic solutions analytically in Appendix \ref{sec:app_B}. In Table \ref{tab:1} we summarize all asymptotic states of branches I and II.

\begin{table}[h!]
\centering
\begin{tabular}{||c c c c c c||}
\hline 
 & & & & & \\ [-1ex]
\, Branch \, & \phantom{gggg} $\phi'$ \phantom{gggg} & \phantom{gggg} $H$ \phantom{gggg} & \phantom{gggg} $a$ \phantom{gggg} & \phantom{gggg} Initial Condition \phantom{gggg} & \, Description \, \\  [1ex] 
\hline
 & & & & & \\
 & $\tau$ & $\tau^{-1}$ & $\tau$ & $\lambda^{3}\phi_{0} = -\Lambda$, $\phi' \to 0^{+}$ & Tadpole--Milne Solution \\
 & & & & & \\
I & $(\tau_{s} - \tau)^{-1}$ & $(\tau_{s} - \tau)^{-1}$ & $\log[1 - \tau/\tau_{s}]$ & $\lambda^{3}\phi_{0} < -\Lambda$, $\phi' \to 0^{+}$ & Finite-Time Instability \\
 & & & & & \\
 & $\tau^{-1/3}$ & $\tau^{1/3}$ & $\exp[\tau^{4/3}]$ & $\lambda^{3}\phi_{0} > -\Lambda$, $\phi' \to 0^{+}$ & Eternal Exponential Growth \, \\
 & & & & & \\
\hline 
 & & & & & \\
II & $ -\tau$ & $\tau^{-3}$ & ${\rm const}$ & $\phi' \to 0^{-}$ & Static, Spatially Non-Flat \\  
 & & & & & \\
\hline  
\end{tabular}
\caption{\label{tab:1} The $\tau \to \infty$, or $\tau \sim {\cal O}(\tau_{s})$, asymptotic behaviour of the dynamical fields for the two branches, and the conditions under which the asymptotic state is approached. The solutions are presented analytically in Appendix \ref{sec:app_B}.}
\end{table}

The asymptotic solution of branch II is a variation of the Minkowski space solution found in \cite{Appleby:2020dko}. Specifically, for $\tau \to \infty$ the scalar field takes the following form 

\begin{equation} \phi \to \psi_{0} + \psi_{1}\tau - {\lambda^{3}\tau^{2} \over 2\epsilon} , 
\end{equation} 

\noindent with constant $\psi_{0}, \psi_{1}$, and the scale factor approaches a constant value $a \to \alpha_{0}$. The Friedmann equation in the same limit reads 

\begin{equation}
\Lambda + \dfrac{M_{\rm pl}^2}{\alpha_0^2} + \lambda^3 \psi_0 + \dfrac{\epsilon \psi_1^2}{2} = 0 \,.
\end{equation}

\noindent This is very similar to the self-tuning solution as reviewed in Section \ref{sec:WTMS} for flat Minkowski space, but with an additional term $M_{\rm pl}^{2}/\alpha_{0}^{2}$ due to the intrinsic curvature of the spatial hypersurfaces. This source of spacetime curvature is also screened, in the $\tau \to \infty$ asymptotic limit. The value of $\alpha_{0}$ depends on the dynamics leading up to the $\tau \to \infty$ solution, and does not approach a common attractor value.

\section{Generalization to FLRW Spacetimes}
\label{sec:generalizationtoflrw}

In Section \ref{sec:cubicgalileonandmilne} we have shown that exact solutions such as Milne can exist even without degeneracy, but that they belong to a more general family characterized by the initial conditions of the scalar field. The Milne solution can be considered as an FLRW spacetime containing a perfect fluid with equation of state $w = -1/3$. One can therefore speculate that similar self-tuning solutions may exist in the presence of other perfect fluids. In this section, we consider the flat FLRW metric 

\begin{equation} ds^{2} = -dt^{2} + a(t)^{2} \left( dr^{2} + r^{2} d\Omega^{2}\right) , 
\end{equation} 

\noindent and assume the existence of an exact solution to the field equations of the form

\begin{equation} -2 M_{\rm pl}^{2} \dot{H} = \left(1 + w_{\rm m} \right) \rho_{\rm m} , \end{equation} 

\noindent with $3M_{\rm pl}^{2} H^{2} = \rho_{\rm m}$ that is, $H = 2/[3(1+w_{\rm m})t]$. This spacetime contains some perfect fluid energy density with constant equation of state $w_{\rm m} = P_{\rm m}/\rho_{\rm m}$. Anticipating the result, we take the Lagrangian (\ref{eq:genm}) and fix $G_{3}(X) = -\epsilon^{2}g_{0}X/\lambda^{3}$, $K(X) = \epsilon X$, with constant dimensionless parameters $g_{0}$, $\epsilon$. With this ansatz, we can write the field equations as 

\begin{eqnarray} \label{eq:hub2} & & 0  = \Lambda + \epsilon {\dot{\phi}^{2} \over 2} - {2 \epsilon^{2} g_{0} \dot{\phi}^{3} \over \lambda^{3}(1+w_{\rm m})t}   + \lambda^{3} \phi , \\ 
\label{eq:dh1} & & 0 =   {\epsilon^{2}g_{0} \over \lambda^{3}}\dot{\phi}^{2}\ddot{\phi} - {2 \over (1+w_{\rm m})t} {\epsilon^{2} g_{0} \over \lambda^{3}}\dot{\phi}^{3}  +  \epsilon \dot{\phi}^{2}   ,  \\
\label{eq:sfe1} & & 0 =   \left[ \epsilon - {4 \epsilon^{2} g_{0} \dot{\phi} \over \lambda^{3}(1+w_{\rm m})t} \right] \ddot{\phi} + {2 \epsilon \over (1+w_{\rm m})t}\dot{\phi}  + \lambda^{3} - {2 \epsilon^{2} g_{0} (1 - w_{\rm m})  \over \lambda^{3}(1 + w_{\rm m})^{2}t^{2}} \dot{\phi}^{2} \,.
\end{eqnarray} 

\noindent We are in a situation analogous to Section \ref{sec:cubicgalileonandmilne} in which we applied the Milne metric \eqref{eq:milne} as an ansatz to the Einstein and scalar field equations and found a solution despite the system being over-constrained. If we select $g_{0}$ such that 

\begin{equation}\label{eq:g0} g_{0} = \left[(1+w_{\rm m})(3+w_{\rm m}) \over (1-w_{\rm m})^{2} \right] ,
\end{equation} 

\noindent assuming $w_{\rm m} \neq 1, -1, -3$, then there exists an exact solution to this system of equations,

\begin{equation} \phi(t) = \phi_{0} + {\lambda^{3} (1 - w_{\rm m}) \over 2 \epsilon(3+w_{\rm m})} t^{2} \,,
\end{equation} 

\noindent together with $\Lambda + \lambda^3 \phi_0 = 0$. We interpret this result such that for any $g_{0}$, there is a power law cosmological model $H \propto t^{-1}$ with associated energy density $\rho_{\rm m}$ for which the vacuum energy is perfectly screened. For example if $\epsilon = 1$ and $g_{0} = 3$, then for a dust filled spacetime there exists an exact matter epoch solution for which $H = 2/[3t]$. That does not mean that $g_{0}$ must take precisely this value to ensure a viable matter epoch -- we have seen that models of this type can admit approximate matter epochs for $g_{0} \neq 3$. The point is, there exists families of perfect fluid self-tuned solutions as we allow the cubic Galileon coefficient to vary. We are not free to specify $\rho_{\rm m}$, $w_{\rm m}$ -- these correspond to the matter content of the Universe. We simply highlight the existence of a family of screened solutions as we vary $g_{0}$. 

The degenerate Minkowski space solution found in \cite{Appleby:2020dko} is given by $g_{0} = 1$, which can be considered as the $w_{\rm m} \to \infty$ limit in this section. The Milne solution in Section \ref{sec:cubicgalileonandmilne} also belongs to this family, because curvature acts as a perfect fluid with equation of state $w_{\rm m} = -1/3$. The value $w_{\rm m} = -1/3$, also corresponds to $g_{0} = 1$, confirming that this particular case admits two distinct scalar-field foliations of Minkowski space. 

The presence of matter generically generates a curvature singularity at $t=0$, with certain special exceptions. In the Milne case, we expanded the scale factor and scalar field around $\tau=0$, but the scale factor is non-analytic around this point for $w_{\rm m} \neq -1/3$. Rather than expanding $a(t)$, we write the general field equations -- including an arbitrary fluid with constant equation of state $w_{\rm m}$ and free parameter $g_{0}$ -- as 

\begin{eqnarray}
 & & 3M_{\rm pl}^{2}H^{2}=\rho_{\rm m} + {\epsilon \over 2} \dot{\phi}^{2} + \lambda^{3}\phi + \Lambda - {3 \epsilon^{2} g_{0} \over \lambda^{3}}H \dot{\phi}^{3}  ,  \\ 
 & & -2M_{\rm pl}^{2} \dot{H} = (1+w_{\rm m})\rho_{\rm m} + \epsilon\dot{\phi}^{2} + {\epsilon^{2}g_{0} \over \lambda^{3}}\dot{\phi}^{2}\ddot{\phi} - {3 \epsilon^{2}g_{0} \over \lambda^{3}}H\dot{\phi}^{3} , \\
 & & \left( \epsilon - {6\epsilon^{2}g_{0} \over \lambda^{3}}H\dot{\phi}\right) \ddot{\phi}  + 3 \epsilon H \dot{\phi} + \lambda^{3} - {9\epsilon^{2} g_{0} \over \lambda^{3}}H^{2}\dot{\phi}^{2} - {3 \epsilon^{2}g_{0} \over \lambda^{3}}\dot{H}\dot{\phi}^{2} , \\
 & & \dot{\rho}_{\rm m} + 3(1+w_{\rm m})H \rho_{\rm m} = 0 ,
\end{eqnarray}

\noindent where we have not imposed any relation between $g_{0}$ and $w_{0}$. We can construct a singular expansion of the variables in this system -- $\rho_{\rm m}$, $H$ and $\phi$. Specifically, we adopt the following singular expansion about $t=0$;

\begin{eqnarray} & & H = {h_{-1} \over t} + \sum_{n=0}^{\infty} h_{n}{t^{n} \over n!} \\ 
& & \rho_{\rm m} = {\rho_{-2} \over t^{2}} + {\rho_{-1} \over t} +  \sum_{n=0}^{\infty} \rho_{n}{t^{n} \over n!} \\ 
& & \phi = \sum_{n=0}^{\infty} \phi_{n}{t^{n} \over n!}
\end{eqnarray} 

\noindent Inserting these into the field equations yields the following impositions on the parameters; $h_{-1} = 2/[3(1+w_{\rm m})]$, $\rho_{-2} = 3h_{-1}^{2}M_{\rm pl}^{2}$, $\phi_{1}=0$, $\rho_{-1} = 0$, $h_{0} = 0$ and $\rho_{0} = -2h_{1} M_{\rm pl}^{2}/(1+w_{\rm m})$. The important $\phi_{2}$ term is given by  

\begin{equation} \phi_{2} = {\lambda^{3} (1+w_{\rm m}) \left(3 + w_{\rm m} \pm \sqrt{(3+w_{\rm m})(3+8g_{0} + w_{\rm m})} \right) \over 4\epsilon g_{0}(3+w_{\rm m})} ,
\end{equation} 

\noindent and the Friedmann equation, at the first non-singular order, reads 

\begin{equation} \lambda^{3}\phi_{0} + \Lambda - {6M_{\rm pl}^{2} h_{1} \over 1+w_{\rm m}} = 0 \, . \end{equation}

\noindent This expansion bears some similarity to the Milne $\tau = 0$ expansion -- the singular nature of the $t=0$ point fixes $\phi_{1} = 0$, but there is a remaining free parameter $\phi_{0}$. If we fix $g_{0}$ according to the expression (\ref{eq:g0}) and also $\lambda^{3}\phi_{0} + \Lambda = 0$, then the power law $H=2/[3(1+w_{\rm m})t]$ is an exact solution, with $h_{1} = 0$, $\rho_{0} = 0$ and all higher order terms zero. However, any other choice of $\phi_{0}$ or $g_{0}$ introduces modified dynamics with potentially $h_{1}$ or higher order terms being non-zero, leading to an initial epoch of perfect-fluid-like domination, transitioning to some modified dynamics in the asymptotic future. We stress that the perfect-fluid-dominating initial phase is present regardless of the values of $g_{0}$, $\phi_{0}$. These constants determine how long the phase persists before the dynamics are affected by the scalar field and vacuum energy. The solution space has two branches characterized by $\phi_{2}$, similarly to Milne. In fact the Milne result can be regarded as a subset of this more general space of solutions. Note the significant difference however; in the Milne solution the $\tau = 0$ point is a coordinate artifact, whereas in the presence of matter the $t=0$ point is a curvature singularity.

\section{Other Milne Solutions}
\label{sec:othermilnesolutions}

We have shown that the specific Lagrangian (\ref{eq:wt}) admits a Milne solution. In this section we briefly present a second model that also admits an exact Milne spacetime, to show that (\ref{eq:wt}) is not unique. Starting from the Lagrangian

\begin{equation}\label{eq:genm} {\cal L} = {M_{\rm pl}^{2} \over 2} R + K(X) - G_{3}(X) \Box \phi - \lambda^{3}\phi - \Lambda   ,
\end{equation} 

\noindent then certain $K(X)$ and $G_{3}(X)$ functions admit exact Milne solutions. For example, taking the particular functions 

\begin{eqnarray} & & K(X) = \alpha X^{3/2} ,  \\
& & G_{3}(X) = \beta X^{2} ,
\end{eqnarray}

\noindent then if the mass scales are related according to $\beta = -9\alpha^{2}/(25\lambda^{3})$, there exists an exact solution to the field equations of the form 

\begin{equation} \phi(\tau) = \phi_{0} + \phi_{2}\tau^{3/2} , \end{equation} 

\noindent where $\Lambda + \lambda^{3}\phi_{0} = 0$, $a(\tau) = \tau$, and the constant $\phi_{2}$ is given by $\phi_{2}^{4} = -8\lambda^{3}/(81\beta)$. This solution cannot be extended to $\tau < 0$ due to the presence of fractional $\tau$ powers, and $\phi''$ diverges at $\tau = 0$.

Unfortunately, finding the general Horndeski model space that admits Milne solutions is not straightforward. For such a solution to exist, there must be a mapping between $\tau$ and $\dot{\phi}$, but we have not found an algorithmic method to construct the mapping in the general case. The situation is more complicated than the Well-Tempering class, for which any model satisfying the condition 

\begin{equation} G_{3X} = -{1 \over \lambda^{3}}K_{X}\left( K_{X} + 2XK_{XX} \right) \end{equation} 

\noindent admits an exact Minkowski space solution, where $X$ subscripts denote derivatives with respect to the argument $X$.

\section{Discussion} 
\label{sec:discussion}

We have shown that the tadpole--cubic Galileon Lagrangian \eqref{eq:wt} accommodates an exact Milne solution that is distinguished by a new kind of self tuning that relies on partially overlapping solution spaces to the non-linear equations of motion, rather than through degeneracy. We have explored the dynamics of this solution, finding that it is not an attractor, but rather that it belongs to a more general family of solutions characterised by the initial condition of the scalar field. By varying the initial field configuration, we find a common Milne-like epoch with $a(\tau) \simeq \tau$ around $\tau \simeq 0$, with the scale factor evolving towards four different final states given in Table \ref{tab:1}. The fact that the vacuum state is not an attractor sets this class apart from Well Tempering \cite{Appleby:2020dko} and the Fab Four \cite{Charmousis:2011bf,Charmousis:2011ea}, and means that the vacuum solution will not persist through a phase transition. Particularly interesting is the existence of two branches of solutions characterised by the scalar field initial velocity $\phi' \to 0^{\pm}$ as $\tau \to 0$, which raises the question of whether the spacetime is extendible to $\tau < 0$. 

When perfect fluids are introduced and a flat FLRW metric ansatz is implemented, the point $t =0$ is a curvature singularity. The dynamical equations do admit a singular expansion around $t =0$, and an analogous situation develops to Milne; two branches of solutions characterized by the field value $\phi_{0}$ at $t=0$, a requirement that $\dot{\phi} =0$ at $t=0$, and an approximate epoch of perfect fluid domination. The late time asymptotic states will depart from GR, in a way that can be tested with data \cite{Bernardo:2022vlj,Escamilla-Rivera:2023rop}. The issue of non-uniqueness of solutions at $t=0$ is overshadowed by the bigger problem of divergences in the curvature invariants. 

The generic methodology of self tuning considered in this work involves a dynamical cancellation of the vacuum energy, which requires that at least one field does not relax to a constant vacuum expectation value. This shifts the CC problem into an initial condition problem. When other energy densities are introduced and the dynamics are `off-shell', the scalar field and scale factor evolve in a way that departs from the standard cosmology. This is true for Fab Four, Well Tempering and the Galileon-Milne system considered in this work. The vacuum energy does affect the expansion rate of  spacetime, but not in the standard way and specifically de Sitter solutions of the form $H^{2} \propto \Lambda$ are not generically present.

The initial state of the Universe is normally considered a high energy problem -- late time cosmology can be studied without recourse to Planck scale physics. Our analysis describes the classical evolution of spacetime and a scalar field, and we have argued that self-tuning mechanisms are dependent on initial conditions. Can we therefore say anything about the cosmological constant problem, or have we pushed the issue back into the realm of high energy physics? Instead of the standard question; `is there is some UV completion of GR that yields a naturally small vacuum energy?' we have arrived at the equally irresolute `is there is some UV completion of gravity that yields a Horndeski scalar-tensor theory that dynamically cancels the CC?' Does this correspond to any progress at all? One interesting avenue is that these models could potentially regularize singular initial conditions. This could move the initial condition problem into the purview of classical or semi-classical physics. Throughout this work we have made the choice $a(\tau =0) = 0$, which is a unique initial state. Other initial conditions, in which the metric is regular or the singularity is a coordinate artifact (as in Milne), can be explored. Based on CMB observations, we know that high energy inflationary models are favoured in the early Universe. Embedding a self tuning model of the CC into the inflationary paradigm is the long term goal. Alternatively, this entire class of models could be constrained or ruled out by local tests of gravity if one could derive the weak field limit. This is an on-going consideration, complicated by the fact that the metric will not be static. Rather, spacetime curvature will be simultaneously induced by the time dependent scalar field and the presence of an inhomogeneous source term. 

\acknowledgements
SA is supported by an appointment to the JRG Program at the APCTP through the Science and Technology Promotion Fund and Lottery Fund of the Korean Government, and were also supported by the Korean Local Governments in Gyeongsangbuk-do Province and Pohang City. We thank Eric Linder for helpful comments on an earlier version of the draft. We also thank Jan Tristram Acu\~{n}a and Che-Yu Chen for important exchanges over the BH analogy of Milne and the finite time instability.

\appendix

\section{Milne and the black hole interior}
\label{sec:analogy}

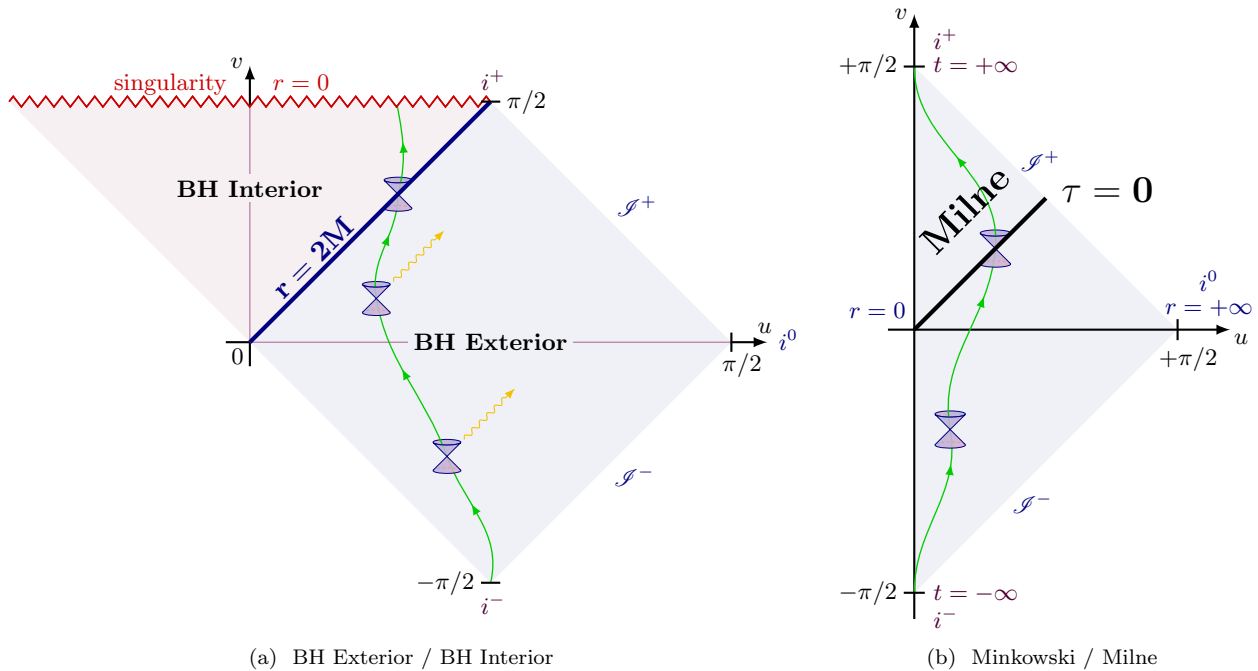
\begin{figure}[h!]
    \centering
	\subfigure[ \ \ {BH Exterior / BH Interior} ]{
        \begin{tikzpicture}[scale=3.2]
          \message{Extended Penrose diagram: Schwarzschild black hole^^J}
          
          \def\R{0.08} 
          \def\Nlines{3} 
          \pgfmathsetmacro\ta{1/sin(90*1/(\Nlines+1))} 
          \pgfmathsetmacro\tb{sin(90*2/(\Nlines+1))}   
          \pgfmathsetmacro\tc{1/sin(90*2/(\Nlines+1))} 
          \pgfmathsetmacro\td{sin(90*1/(\Nlines+1))}   
          \coordinate (-O) at (-1, 0); 
          \coordinate (-N) at (-1, 1); 
          \coordinate (O)  at ( 1, 0); 
          \coordinate (S)  at ( 1,-1); 
          \coordinate (N)  at ( 1, 1); 
          \coordinate (E)  at ( 2, 0); 
          \coordinate (W)  at ( 0, 0); 
          \coordinate (B)  at ( 0,-1); 
          \coordinate (X0) at ({asin(sqrt((\ta^2-1)/(\ta^2-\tb^2)))/90},
                               {-acos(\ta*sqrt((1-\tb^2)/(\ta^2-\tb^2)))/90}); 
          \coordinate (X1) at ({asin(sqrt((\tc^2-1)/(\tc^2-\td^2)))/90},
                               {acos(\tc*sqrt((1-\td^2)/(\tc^2-\td^2)))/90}); 
          \coordinate (X2) at (45:0.87); 
          \coordinate (X3) at (0.60,1.05); 
          
          \draw[->,thick] (0,-0.1) -- (0,1.15) node[above=1,left=-1] {$v$};
          \draw[->,thick] (-0.1,0) -- (2.15,0) node[left=1,above=0] {$u$};
          
          \begin{scope}
            
            \clip[decorate,decoration={zigzag,amplitude=2,segment length=6.17}]
              (-N) -- (N) --++ (1.1,0.1) |-++ (-3.1,-2.3) -- cycle;
            
            \fill[mylightpurple] (-N) |-++ (2,0.1) -- (N) -- (W) -- cycle;
            \fill[mylightblue] (N) -- (E) -- (S) -- (W) -- cycle;
            
            \coneback{X0};
            \coneback{X1};
            \coneback{X2};
            
            \draw[world line t] (W) -- (E) (W) -- (0,1.1);
            
            \draw[particle,decoration={markings,mark=at position 0.16 with {\arrow{latex}},
                                                mark=at position 0.45 with {\arrow{latex}},
                                                mark=at position 0.72 with {\arrow{latex}},
                                                mark=at position 0.90 with {\arrow{latex}}},postaction={decorate}]
              (S) to[out=77,in=-70] (X0) to[out=110,in=-80] (X1)
                  to[out=100,in=-90] (X2) to[out=75,in=-80] (X3);
          
          \end{scope}
          
          \conefront{X0};
          \conefront{X1};
          \conefront{X2};
          
          \draw[photon] (X0) ++ (45:0.1) --++ (45:0.3);
          \draw[photon] (X1) ++ (45:0.1) --++ (45:0.3);
          
          \node[fill=mylightblue,inner sep=2, font=\bfseries] at (O) {BH Exterior};
          \node[fill=mylightpurple,inner sep=2, font=\bfseries] at (0,0.64) {BH Interior};
          
          \draw[singularity] (-N) -- node[pos=0.46,above left=-2] {\strut singularity} (N);
          \draw[singularity] (-N) -- node[pos=0.54,above right=-2] {\strut $r=0$} (N);
        
          \path (W) -- (N) node[mydarkblue,pos=0.32,above=-2.5,rotate=45,scale=1.2,font=\bfseries]
          {\contour{mylightpurple}{{$\mathbf{r = 2M}$}}};
        \draw[ultra thick,mydarkblue] (W) -- (N);

          \node[below left=-1] at (W) {$0$};
          \tick{E}{90} node[right=4,below=-3] {$\pi/2$};
          \tick{S}{0} node[left=-1] {$-\pi/2$};
          \tick{N}{180} node[right=-1] {$\pi/2$};
          
          \node[above=1,right=1,mydarkblue] at (2.15,0) {$i^0$};
          \node[right=1,below=1,mydarkpurple] at (S) {$i^-$};
          \node[right=1,above=1,mydarkpurple] at (N) {$i^+$};
          \node[mydarkblue,above right=-1] at (1.5,0.5) {$\calI^+$};
          \node[mydarkblue,below right=-2] at (1.5,-0.5) {$\calI^-$};
          
        \end{tikzpicture}
    }
    \subfigure[\ \ {Minkowski / Milne} ]{
        \begin{tikzpicture}[scale=3.5]
          \message{Penrose diagram (radius r)^^J}
          
          \def\Nlines{4} 
          \def\ta{tan(90*1.0/(\Nlines+1))} 
          \def\tb{tan(90*2.0/(\Nlines+1))} 
          \coordinate (O) at ( 0, 0); 
          \coordinate (S) at ( 0,-1); 
          \coordinate (N) at ( 0, 1); 
          \coordinate (E) at ( 1, 0); 
          \coordinate (X) at ({penroseu(\tb,\tb)},{penrosev(\tb,\tb)});
          \coordinate (X0) at ({penroseu(\ta,-\tb)},{penrosev(\ta,-\tb)});
          
          \fill[mylightblue] (N) -- (E) -- (S) -- cycle;
          \draw[->,thick] (-0.1,0) -- (1.2,0) node[below right=-2] {$u$};
          \draw[->,thick] (0,-1.1) -- (0,1.2) node[left=-1] {$v$};
          
          \node[above=1,above left=0,mydarkblue,align=center] at (O)
            {$r=0$};
          \node[left=6,above right=-2,mydarkblue,align=center] at (1,0.04)
            {$i^0$\\[-2]$r=+\infty$};
          \node[above=6,below right=0,mydarkpurple,align=left] at (0.04,-1)
            {$t=-\infty$\\$i^-$};
          \node[below=6,above right=0,mydarkpurple,align=left] at (0.04,1)
            {$i^+$\\[-2]$t=+\infty$};
          \node[mydarkblue,above right,align=right] at (57:0.68)
            {$\calI^+$};
          \node[mydarkblue,below right,align=right] at (-60:0.68)
            {$\calI^-$};
          
          \coneback{X};
          \coneback{X0};
        
          \draw[particle,decoration={markings,mark=at position 0.24 with {\arrow{latex}},
                                              mark=at position 0.55 with {\arrow{latex}},
                                              mark=at position 0.82 with {\arrow{latex}}},postaction={decorate}]
            (S) to[out=90,in=-80] (X0) to[out=100,in=-95] (X) to[out=85,in=-90] (N);
          
          \conefront{X};
          \conefront{X0};
        
          \draw[ultra thick, black] (O) -- (0.5,0.5) node[above=3,right=2,font=\Large\bfseries] {$\mathbf{\tau=0}$};
          \node[above, font=\Large\bfseries, rotate=45] at (0.25, 0.4) {Milne};

          \tick{E}{90} node[right=4,below=-1] {$+\pi/2$};
          \tick{S}{ 0} node[left=-1] {$-\pi/2$};
          \tick{N}{ 0} node[left=-1] {$+\pi/2$};
          
        \end{tikzpicture}
    }
    \caption{{Conformal diagrams of the Schwarzschild black hole and the Minkowski/Milne spacetime.
    }}
    \label{fig:conformaldiagrams}
\end{figure}

An analogy between Milne/Minkowski space and the black hole (BH) interior can be established by referring to conformal diagrams \cite{2007EJPh...28..877S} (Figure \ref{fig:conformaldiagrams}). This illustrates the horizon-like role of $\tau = 0$ causal boundary that separates the Milne patch of Minkowski.

The conformal diagram of a Schwarzschild black hole reveals that the event horizon at $r = 2M$ acts as a one-way membrane that causally disconnects the BH interior from the exterior region. Light cones drawn over arbitrary points inside the BH show that the horizon lies outside the future light cone of any interior point. As a result, observers in the BH exterior cannot receive signals from the interior, nor can observers inside the black hole access the horizon by sending signals. This is independent of coordinate choice by causality preservation.

An analogous situation comes up in Minkowski space and its Milne patch. In Minkowski, all points can be reached by time and spatial translations. However, when considering its Milne patch, the $\tau = 0$ boundary acts analogously to a horizon in a BH spacetime such that the Milne patch becomes causally disjoint from the non-Milne volume of the full Minkowski space \cite{Castorina:2007eb}. Being a special point in the Milne coordinate system, $\tau = 0$ thus acts to artificially disinherit the time translation invariance of Minkowski. Furthermore, observers in the non-Milne volume cannot obtain signals from the Milne patch, and likewise  between observers in the Milne patch and the $\tau = 0$ surface. In this sense, the $\tau = 0$ Milne boundary acts as a hideout of events just as an event horizon would for a black hole.

\section{Non-Equivalence of the field equations \eqref{eq:dhm1} and \eqref{eq:sfem}}
\label{sec:nonlinearity}

In this appendix we explicitly show that the dynamical equations \eqref{eq:dhm1} and \eqref{eq:sfem} are not equivalent, by presenting the complete set of solutions to \eqref{eq:dhm1} and then showing that the general solution does not solve \eqref{eq:sfem}. Only a subset simultaneously solves both equations. This implies that degeneracy is not realised.

Because the equation factorizes, all solutions of \eqref{eq:dhm1} can be obtained. They are given by
\begin{equation}
    \phi(\tau) =
    \begin{cases}
        \overline{c}_1 & \,,\,\,\,\, {\rm constant} \\
        c_1 + \dfrac{c_2 \tau^4}{4} + \dfrac{\lambda^3 \tau^2}{4 \epsilon} & \,,\,\,\,\, {\rm time \ dependent}
    \end{cases}
\end{equation}
where $c_1, c_2$ and $\overline{c}_1$ are integration constants. The substitution of the constant solution into \eqref{eq:sfem} leads to $\lambda = 0$ for any $\overline{c}_1$. Since we are taking $\lambda \neq 0$, this shows that the constant solution of \eqref{eq:dhm1} is not a solution in \eqref{eq:sfem}. 

Substitution of the time dependent solution into \eqref{eq:sfem} leads to the condition $c_2 = 0$, whereas any value of the integration constant $c_2 \neq 0$ solves \eqref{eq:dhm1}. The time dependent solution of \eqref{eq:dhm1} is therefore generally not permissible to \eqref{eq:sfem}, except for the special case $c_2 = 0$ which happens to be the Milne solution \eqref{eq:mils}. Hence we have inferred that \eqref{eq:dhm1} and \eqref{eq:sfem} are independent equations by showing that they have only partially overlapping solution spaces. It is also possible to show that the solutions of \eqref{eq:sfem} are not acceptable to \eqref{eq:dhm1}, except for the Milne subset, leading to the same conclusion.  

\section{Alternative Coordinate Representations of Minkowski Space Solutions}
\label{sec:alternativeminkowski}

The Lagrangian given in (\ref{eq:wt}) contains exact Minkowski space and Milne solutions. These are distinct, not related via a coordinate transformation. To manifest this distinction, in this appendix we derive the Milne solution in Minkowski coordinates and vice versa. One could instead perform a coordinate transformation on the solution and arrive at the same result.

\subsection{Minkowski Solution in Milne Coordinates} 
\label{sec:Mink_in_Milne} 

If we use the metric (\ref{eq:gen_milne}) and the ansatz $\phi = \phi(\tau,y)$, the gravitational and scalar field equations read;

\begin{equation}
\label{eq:ap1}
\begin{split}
    3 M_{\rm pl}^2 H^2 = \ & \lambda ^3 \phi + \Lambda -\dfrac{\epsilon ^2 \left(k y^2-1\right) \phi_y^2}{\lambda ^3 a^4} \left(\left(k y^2-1\right) \phi_{yy} + k y \phi_y\right) -\frac{1}{2} \epsilon  \left(\frac{\left(k y^2-1\right) \phi_y^2}{a^2}+\phi_\tau^2\right) + \epsilon  \phi_\tau^2 \\
    & \ \ - \frac{\epsilon ^2 H \left(k y^2-1\right)}{\lambda ^3 a^2} \phi_y^2 \phi_\tau - \frac{\epsilon ^2 \left(3 k y^2-2\right)}{\lambda ^3 y a^2} \phi_y \phi_\tau^2 - \frac{\epsilon ^2 \left(k y^2-1\right)}{\lambda ^3 a^2} \phi_{yy} \phi_\tau^2 -\frac{3 \epsilon ^2 H \phi _\tau^3}{\lambda ^3} - \dfrac{3 k M_{\rm pl}^2}{a^2}
\end{split}
\end{equation}
\begin{equation}
\label{eq:ap2}
\begin{split}
    2 M_{\rm pl}^2 H' = \ &-\frac{\epsilon ^2 \left(k^2 y^4-3 k y^2+2\right)}{\lambda ^3 y a^4}\phi_y^3-\frac{\epsilon ^2 \left(2-3 k y^2\right)}{\lambda ^3 y a^2}\phi_\tau^2 \phi_y +\frac{3 \epsilon ^2 H }{\lambda ^3} \phi _\tau^3  +\frac{2 k M_{\rm pl}^2}{a^2} \\
    & \ \ + \phi_\tau^2 \left(-\frac{\epsilon ^2 \left(1-k y^2\right) }{\lambda ^3 a^2} \phi_{yy} -\frac{\epsilon ^2 }{\lambda ^3} \phi_{\tau\tau} - \epsilon \right) + \phi_y^2 \left(k y^2-1\right) \bigg[-\frac{\epsilon ^2 H }{\lambda ^3 a^2}\phi_\tau + \frac{\epsilon ^2 \left(k y^2-1\right) }{\lambda ^3 a^4} \phi_{yy} - \frac{\epsilon ^2 }{\lambda ^3 a^2} \phi_{\tau\tau} +\frac{\epsilon }{a^2}\bigg]
\end{split}
\end{equation}
\begin{equation}
\label{eq:ap3}
\begin{split}
    0 = \ & \lambda ^3-\frac{\epsilon ^2 \left(k y^2-1\right)}{\lambda ^3 y^2 a^4} \phi_y^2 \left(y^2 a^2 \left(H' + H^2 \right) +8 k y^2-2\right) + \frac{2 \epsilon ^2 \left(k y^2-1\right)}{\lambda ^3 a^2} \phi_{\tau y}^2 +\phi_\tau^2 \left(-\frac{3 \epsilon ^2 H'}{\lambda ^3}-\frac{9 \epsilon ^2 H^2}{\lambda ^3}\right) \\
    & \ \ + \phi_{yy} \left(\frac{\epsilon  \left(k y^2-1\right)}{a^2}-\frac{2 \epsilon ^2 \left(k y^2-1\right) }{\lambda ^3 a^2}\phi_{\tau\tau} \right) + \epsilon \phi_{\tau\tau} + \phi_\tau \left(-\frac{4 \epsilon ^2 H\left(k y^2-1\right) }{\lambda ^3 a^2} \phi_{yy} -\frac{6 \epsilon ^2 H }{\lambda ^3} \phi_{\tau\tau} +3 \epsilon  H\right) \\
    & \ \ + \phi_y \bigg[ \frac{\epsilon  \left(3 k y^2-2\right)}{y a^2}-\frac{4 \epsilon ^2 \left(k y^2-1\right)^2 }{\lambda ^3 y a^4} \phi_{yy} -\frac{4 \epsilon ^2 H \left(3 k y^2-2\right) }{\lambda ^3 y a^2} \phi_\tau - \frac{4 \epsilon ^2 H\left(k y^2-1\right) }{\lambda ^3 a^2} \phi_{\tau y} -\frac{2 \epsilon ^2 \left(3 k y^2-2\right) }{\lambda ^3 y a^2}\phi_{\tau\tau} \bigg] \,,
\end{split}
\end{equation}
where subscripts $\tau$ and $y$ on $\phi=\phi(\tau, y)$ imply partial differentiation with respect to that variable, e.g., $\phi_{yy} = \partial_y^2 \phi$ and $\phi_{\tau y} = \partial_\tau \partial_y \phi$.
The degenerate Minkowski space solution can be recovered by fixing $k = -1$, $a(\tau) = \tau$, $H(\tau) = 1/\tau$ and substituting $\phi(\tau, y) = \phi\left( \tau \sqrt{1 + y^2} \right)$. In doing so, equations (\ref{eq:ap2},\ref{eq:ap3}) both reduce to $\epsilon\ddot{\phi}\left(\tau \sqrt{1 + y^2}\right) + \lambda^3= 0$ where dots correspond to differentiation over the argument $t = \tau \sqrt{1+ y^2}$. This relation is exactly the coordinate transformation between Milne and Minkowski coordinates. The dynamical equations can thus be identified as \eqref{eq:wt2} and \eqref{eq:wt3} which then gives rise to the degenerate solution.

When the scalar field and/or metric components possess explicit spatial dependence, the momentum constraint equation is no longer trivially satisfied. For the metric (\ref{eq:gen_milne}) and ansatz $\phi(y,\tau)$, the $y-\tau$ component of the Einstein equations reads --

\begin{equation}\label{eq:ytau}
\begin{split}
    0 = \epsilon  y H \phi_y \left(3 a^2 \phi_\tau^2+\left(k y^2-1\right) \phi_y^2\right)-y a^2 \phi_\tau \left(\epsilon  \phi_\tau \phi_{\tau y}+\lambda ^3 \phi_y\right)+\epsilon  \left(k y^2-1\right) \phi_y^2 \left(2 \phi_\tau-y \phi_{\tau y}\right) \,.
\end{split}
\end{equation}

\noindent If we impose $a(\tau) = \tau$, $k=-1$ and $\phi = \phi\left(\tau\sqrt{1+y^{2}}\right)$, then (\ref{eq:ytau}) also reduces to $\epsilon\ddot{\phi}\left(\tau \sqrt{1 + y^2}\right) + \lambda^3= 0$.

\subsection{Milne Solution in Minkowski Coordinates} 
\label{sec:Milne_in_Mink}

Similarly, the Milne solution can be recast in Minkowski coordinates. If we take the metric 

\begin{equation} ds^{2} = -dt^{2} + dr^{2} + r^{2}d\Omega^{2}  
\end{equation} 

\noindent and search for solutions to the field equations of the form $\phi = \phi(t, r)$, the equations read ;

\begin{eqnarray}\label{eq:app_a1} & & \lambda^{3}\phi + \Lambda - {\epsilon^{2} \over \lambda^{3}}(\phi_{r})^{2}\phi_{rr} - {\epsilon \over 2} \left(\phi_{t}^{2} - \phi_{r}^{2} \right) + \epsilon \phi_{t}^{2}+{2 \over r}{\epsilon^{2} \over \lambda^{3}}\phi_{r}\phi_{t}^{2} + {\epsilon^{2} \over \lambda^{3}}\phi_{t}^{2}\phi_{rr} = 0 \\ 
\label{eq:app_a2} & & -{2 \over r} {\epsilon^{2} \over \lambda^{3}}\phi_{r}^{3} - {2 \over r} {\epsilon^{2} \over \lambda^{3}}\phi_{t}^{2}\phi_{r} - \phi_{t}^{2} \left( {\epsilon^{2} \over \lambda^{3}}\phi_{rr} + {\epsilon^{2} \over \lambda^{3}}\phi_{tt} + \epsilon \right) + \phi_{r}^{2} \left({\epsilon^{2} \over \lambda^{3}}\phi_{rr} + {\epsilon^{2} \over \lambda^{3}}\phi_{tt} - \epsilon \right) = 0 \\
\label{eq:app_a3} & & \lambda^{3} - {2 \over r^{2}}{\epsilon^{2} \over \lambda^{3}}\phi_{r}^{2} - 2 {\epsilon^{2} \over \lambda^{3}} \phi_{rt} + \phi_{r}^{2} \left( -\epsilon + 2 {\epsilon^{2} \over \lambda^{3}}\phi_{tt}\right) + \epsilon \phi_{tt} + \phi_{r}\left( -{2  \over r}\epsilon - {4 \over r} {\epsilon^{2} \over \lambda^{3}}\phi_{rr} + {4 \over r} {\epsilon^{2} \over \lambda^{3}}\phi_{tt} \right) = 0 ,
\end{eqnarray} 

\noindent where $r,t$ subscripts denote derivatives with respect to the $r$ and $t$ coordinates respectively. These equations admit a solution of the form 

\begin{equation} \label{eq:mils_flat} \phi(r,t) = \phi_{0} + {\lambda^{3} \over 4\epsilon} \left(t^{2} - r^{2} \right) , \end{equation} 

\noindent subject to the condition $\lambda^{3}\phi_{0} + \Lambda = 0$, which is the solution to the Friedmann equation (\ref{eq:app_a1}). The result is expected; this is the Milne solution in `flat' Minkowski coordinates. The Milne coordinate patch covers $t \geq  r$, and on the light cone $t=r$ the scalar field is static $\phi = \phi_{0}$.

\section{Analytic Solutions in the large $\tau$ Limit}
\label{sec:app_B}

In the main body of the text we inferred the asymptotic $\tau \to 0$ solutions for Milne-like spacetimes, and then numerically evaluated the equations to find the behaviour of the fields beyond the regime of validity of the expansion. In doing so, we found that branch I possesses two distinct asymptotic states, and branch II a single attractor solution at large $\tau$. In this appendix we derive the analytic behaviour of the dynamics in the large $\tau$ limit, describing each of these final states.

\subsection{Regular Asymptotic $\tau \to \infty$ Solution of Branch I}

There are two asymptotic states for branch I. In this section we derive the exponential $a(\tau)$ solution in the large $\tau$ limit. This was characterised numerically by the initial conditions $\phi' \to 0^{+}$ and $\phi_{0} > -\Lambda/\lambda^{3}$. Noticing that this is a slow-roll state, we can use $H' \ll H^{2}$ and $\phi'' \ll H\phi'$. Then, the scalar field equation reduces to an approximate algebraic relation for the combination $H\phi'$ -- 

\begin{equation} 9{\epsilon^{2} \over \lambda^{3}} H^{2}(\phi')^{2} - 3\epsilon H \phi' - \lambda^{3} \simeq 0 ,
\end{equation} 

\noindent which admits a solution $H\phi' \simeq b \lambda^{3}$, with $b = (1+\sqrt{5})/6\epsilon$. We can then solve the dynamical Einstein equation under the slow roll approximation 

\begin{equation} -2 M_{\rm pl}^{2}H' \simeq \epsilon (\phi')^{2} - 3{\epsilon^{2} \over \lambda^{3}}H (\phi')^{3} \,.
\end{equation}

Using the approximation $H\phi' \simeq b\lambda^{3}$ is constant, multiplying this equation by $H^{2}$ we can find the approximate solution 

\begin{eqnarray} & &  H \simeq \left({b\lambda^{6} \tau \over 2M_{\rm pl}^{2}}\right)^{1/3} \\
& & \phi \simeq \phi_{0} + 3\lambda \left({b M_{\rm pl} \over 2}\right)^{2/3} \tau^{2/3} \\
& & \phi' \simeq 2 \lambda \left({b M_{\rm pl} \over 2}\right)^{2/3} \tau^{-1/3} \\
& & a(\tau) \propto \exp \left[ {3 \over 4}\left({b\lambda^{6} \over 2M_{\rm pl}^{2}}\right)^{1/3}\tau^{4/3} \right] \,.
\end{eqnarray}

The expansion rate $H$ monotonically increases as $\tau \to \infty$ but $H' \to 0$. The solution conforms to the initial ansatz used to derive it; $H' \ll H^{2}$, $\phi'' \ll H \phi'$ as $\tau \to \infty$. The scale factor $a(\tau)$ undergoes perpetual exponential growth; this is not a transitory phase like inflation. Some additional physical process would need to be added to the model to extricate the scale factor from this state.

\subsection{Singular Asymptotic $\tau \to \infty$ Solution of Branch I}

The second asymptotic solution of branch I, which is reached from initial conditions $\phi' \to 0^{+}$ and $\phi_{0} < -\Lambda/\lambda^{3}$, is characterized by an abrupt instability, in which the highest derivatives $H'$ and $\phi''$ grow very rapidly at some finite time. To understand this behaviour, we expand the fields $a$ and $\phi$ about their Milne solutions as 

\begin{eqnarray}
    a(\tau) &=& \tau + \xi \delta a(\tau) \\
    \phi(\tau) &=& \phi_0 + \dfrac{\lambda^3 \tau^2}{4\epsilon} + \xi \delta \phi (\tau) \,.
\end{eqnarray}

\noindent Linearizing the Friedmann constraint, Hubble and scalar field equations with respect to the small parameter $\xi \ll 1$, we find 

\begin{equation}
\label{eq:ab1}
    \left(-\phi_0\lambda ^3-\Lambda \right)+\frac{\xi}{8 \epsilon  \tau ^2} \left(3 \lambda ^6 \tau ^4 \delta a'(\tau )-3 \lambda ^6 \tau ^3 \delta a(\tau )+14 \epsilon  \lambda ^3 \tau ^3 \delta \phi '(\tau )-8 \epsilon  \lambda ^3 \tau ^2 \delta \phi (\tau )+48 \epsilon  \mpl \delta a'(\tau )\right)
 +{\cal O}\left(\xi ^2\right) = 0
\end{equation}
\begin{equation}
\label{eq:ab2}
    \frac{\xi}{8 \epsilon  \tau ^2} \left(-3 \lambda ^6 \tau ^4 \delta a'(\tau )+3 \lambda ^6 \tau ^3 \delta a(\tau )+2 \epsilon  \lambda ^3 \tau ^4 \delta \phi ''(\tau )-6 \epsilon  \lambda ^3 \tau ^3 \delta \phi '(\tau )+16 \epsilon  \tau  M_{\rm pl}^2 \delta a''(\tau )-32 \epsilon  M_{\rm pl}^2 \delta a'(\tau )\right) + {\cal O}\left(\xi ^2\right) = 0
\end{equation}
\begin{equation}
\label{eq:ab3}
    \xi  \left(-\frac{3}{4} \lambda ^3 \tau  \delta a''(\tau )-\frac{3 \left(\lambda ^3 \tau  \delta a'(\tau )+\lambda ^3 (-\delta a(\tau ))+2 \epsilon  \delta \phi '(\tau )\right)}{\tau }-2 \epsilon  \delta \phi ''(\tau )\right)+{\cal O}\left(\xi ^2\right) = 0 \,.
\end{equation}

\noindent We define a `singular' time $\tau_s$, which is the value of $\tau$ at which the coefficient of the highest derivative term $\delta \phi''$ vanishes, as
\begin{equation}
    \tau_s = \left( \dfrac{64 \epsilon \mpl}{3 \lambda^6} \right)^{1/4} \,,
\end{equation}
we obtain the following solution for $\delta a$;
\begin{eqnarray}
    \delta a(\tau) = c_1 + c_2 \tau + c_3 \left( \tau  {\rm arctanh}\left(\frac{\tau ^2}{\tau_s^2}\right)+\frac{\tau_s}{4} \left(\log \left(\frac{\tau_s-\tau }{\tau_s + \tau}\right)+2 \arctan\left(\frac{\tau }{\tau_s}\right)\right) \right) \,.
\end{eqnarray}
Now, by noting the following leading asymptotic behavior of each function above,
\begin{eqnarray}
    \tau {\rm arctanh} \left( 
\dfrac{\tau^2}{\tau_s^2} \right) & \sim & -\dfrac{\tau_s}{2} \ln \left( 1 - \dfrac{\tau}{\tau_s} \right) \\
\arctan \left( \dfrac{\tau}{\tau_2} \right) & \sim & \dfrac{\pi}{4} \\
\ln \left( \dfrac{\tau_s - \tau}{\tau_s + \tau} \right) & \sim & \ln \left( 1 - \dfrac{\tau}{\tau_s} \right) \,,
\end{eqnarray}
we find that the linear perturbation logarithmically diverges at the singular point, $\tau \sim \tau_s$,
\begin{equation}
    \delta a(\tau) \sim c_3 \ln\left( 1 - \dfrac{\tau}{\tau_s} \right) \,. 
\end{equation}

\noindent Next, after integrating the scalar field equation once, we have the following exact relation --

\begin{equation}\label{eq:p1} \epsilon \phi' - 3 {\epsilon^{2} \over \lambda^{3}} H (\phi')^{2} = -a^{-3} \lambda^{3} \int a^{3} d\tau \,.
\end{equation} 

\noindent Linearizing this equation in $a = \tau + \xi \delta a$, and $\phi' = \lambda^{3}\tau/(2\epsilon) + \xi \delta \phi'$, we find a relation between $\delta \phi'$ and $\delta a$;   

\begin{equation} 2\epsilon \delta \phi' + {3 \over 4}\lambda^{3} \tau \delta a' = {3 \lambda^{3} \over \tau^{3}}\int \tau^{2} \delta a d\tau \end{equation} 

\noindent and from this we can determine that the leading order singular term in $\delta \phi'$ on approach to $\tau \to \tau_{s}$ is 

\begin{equation} \delta \phi' \sim -{3 \over 8 \epsilon}\lambda^{3} \tau_{s} \delta a' \sim -{3c_{3} \over 8 \epsilon}\lambda^{3}  {\tau_{s} \over \tau_{s} - \tau} \,.
\end{equation} 

\noindent The linearized approximation used here will break down before the singular point is reached, but the approach to $\tau \to \tau_{s}$ will be well described by the perturbative expansion. The reason why is that the divergence is only logarithmic in the scale factor, and at worst linear $(\tau - \tau_{s})^{-1}$ in $\delta a'$ and $\delta \phi'$. The divergent behaviour is most pronounced in the highest derivatives $\delta a''$ and $\delta \phi''$, and the equations are already linear in these quantities. However, the final stages of the evolution will not be described by this expansion -- the breakdown will occur at the point at which $\delta a' \sim {\cal O}(1)$ and $\delta \phi' \sim {\cal O}(\lambda^{3}\tau/2)$. Beyond this point, the dynamical evolution is inherently non-linear. Regardless, it is unlikely that this sub-branch will yield a viable spacetime that can match the properties of our Universe. 

\subsection{Asymptotic $\tau \to \infty$ Solution of Branch II}

For branch II, which is characterized by the initial condition $\phi' \to 0^{-}$, we have found numerically that there exists an attractor solution as $\tau \to \infty$. We can recover this $\tau \to \infty$ limit by considering the following ansatz

\begin{eqnarray} & & a(\tau) = \alpha_{0} + \delta a(\tau)  \\
& & \phi(\tau) = \psi_{0} + \psi_{1} \tau - {\lambda^{3} \tau^{2} \over 2\epsilon} + \delta \phi(\tau) 
\end{eqnarray} 

\noindent such that $\alpha_{0}$ is a constant, $\delta a$ and $\delta \phi$ approach zero as $\tau \to \infty$, subject to $H \dot{\phi} \to 0$. By expanding in powers of $1/\tau$, we adopt the following form

\begin{eqnarray} & & a(\tau) = \alpha_{0} + \sum_{n=1}^{\infty} \alpha_{-n} \tau^{-n} \\ 
& & \phi(\tau) = \psi_{0} + \psi_{1} \tau - {\lambda^{3} \tau^{2} \over 2\epsilon} + \sum_{n=1}^{\infty} \psi_{-n} \tau^{-n} \,.
\end{eqnarray} 

\noindent Inserting into the field equations, we find $\alpha_{-1} = 0$, $\alpha_{-2} = \epsilon M_{\rm pl}^{2} /(3\lambda^{6}\alpha_{0})$ and the Friedmann equation at lowest order satisfies

\begin{equation}
\Lambda + \dfrac{M_{\rm pl}^2}{\alpha_0^2} + \lambda^3 \psi_0 + \dfrac{\epsilon \psi_1^2}{2} = 0     
\end{equation}

\noindent and at the next order, we have the condition 

\begin{equation} 6\epsilon^{2}M_{\rm pl}^{2} \psi_1 + 2 \alpha_{0}^{2} \epsilon \lambda^{6}\psi_{-1} - 9 \alpha_{0}\lambda^{9}\alpha_{-3} = 0 \,.\end{equation} 

\noindent For this solution, the metric, its first and second derivatives and hence the curvature invariants are finite over the range in which the expansion is valid. This asymptotic solution satisfies $H\dot{\phi} \to \tau^{-3} \to 0$, as required. 

This is not dissimilar to the  $t \to \infty$, asymptotic flat Minkowski solutions found in previous work \citep{Appleby:2022bxp}. The intrinsic curvature of the constant time hypersurfaces contributes a term $3M_{\rm pl}^{2}/\alpha_{0}^{2}$ to the asymptotic Friedmann equation, but beyond that the solution looks similar to the flat Minkowski space solution originally found.


\providecommand{\noopsort}[1]{}\providecommand{\singleletter}[1]{#1}%

\end{document}